\documentclass[%
 aip,pof
 amsmath,amssymb,
 reprint,
]{revtex4-2}

\usepackage{graphicx}
\usepackage{dcolumn}
\usepackage{bm}

\usepackage[utf8]{inputenc}
\usepackage[T1]{fontenc}
\usepackage{mathptmx}
\usepackage{etoolbox}
\usepackage{amssymb}
\usepackage{amsmath}
\usepackage{hyperref}
\usepackage{xcolor}
\definecolor{color}{HTML}{0000FF}
\hypersetup{linkcolor=color,urlcolor=color,citecolor=color,filecolor=color,colorlinks=true}

\makeatletter
\def\@email#1#2{%
 \endgroup
 \patchcmd{\titleblock@produce}
  {\frontmatter@RRAPformat}
  {\frontmatter@RRAPformat{\produce@RRAP{*#1\href{mailto:#2}{#2}}}\frontmatter@RRAPformat}
  {}{}
}%
\makeatother
\begin{document}

\preprint{AIP/123-QED}

\title[Preprint submitted to Physics of Fluids (AIP publishing)]{Nonuniform heating of a substrate in evaporative lithography}
\author{Mohammed~A.~Al-Muzaiqer}
\affiliation{
    Photonics and Microfluidics Lab., X-BIO Institute, University of Tyumen,
     Tyumen 625003, Russia
}
\affiliation{
    Microfiltration Processes Lab., University of Tyumen, Tyumen 625003, Russia
}
\author{Konstantin~S.~Kolegov}
 \email{konstantin.kolegov@asu.edu.ru}
 \homepage{https://science.asu.edu.ru/index.php/user/9\#!profile}
\affiliation{
    Mathematical Modeling Lab., Institute of Physics and Mathematics, Astrakhan State University,  Astrakhan 414056, Russia
}
\affiliation{
    Caspian Institute of Maritime and River Transport, Volga State University of Water Transport, Astrakhan 414000,  Russia
}
\affiliation{
    Landau Institute for Theoretical Physics, Russian Academy of Sciences, Chernogolovka
    142432, Russia
}

\author{Natalia~A.~Ivanova}
\email{n.ivanova@utmn.ru}
\homepage{https://www.utmn.ru/x-bio/about/struktura-instituta/laboratorii/fotonika/}
\affiliation{
    Photonics and Microfluidics Lab., X-BIO Institute, University of Tyumen,
     Tyumen 625003, Russia
}
\affiliation{
    Microfiltration Processes Lab., University of Tyumen, Tyumen 625003, Russia
}

\author{Viktor~M.~Fliagin}
\affiliation{
    Photonics and Microfluidics Lab., X-BIO Institute, University of Tyumen,
     Tyumen 625003, Russia
}
\affiliation{
    Microfiltration Processes Lab., University of Tyumen, Tyumen 625003, Russia
}

\date{\today}

\begin{abstract}
This work is devoted to a method to generate particle cluster assemblies and connected to evaporative lithography. Experiments are carried out using nonuniform evaporation of an isopropanol film containing polystyrene microspheres in a cylindrical cell. The local
inhomogeneity of the vapor flux density is achieved by exploiting  the temperature gradient. A copper rod is mounted in the central part of the bottom of the cell for further heating. The thermocapillary flow resulting from the surface tension gradient, due in turn to the temperature drop, transfers the particles that were originally at rest at the bottom of the cell. The effect of the initial thickness of the liquid layer on the height and base area of the cluster formed in the central region of the cell is studied. The velocity is measured using particle image velocimetry. A model describing the initial stage of the process is developed. The
equations of heat transfer and thermal conductivity are used to
define the temperature distribution in the liquid and in the cell. The fluid flow is simulated using the lubrication approximation. The
particle distribution is modeled using the convection--diffusion
equation. The evaporation flux density is calculated using the
Hertz-Knudsen equation. The dependence of the liquid viscosity on
the particle concentration is described by Mooney's formula.
Numerical results show that the liquid film gradually becomes
thinner in the central region, as the surface tension decreases with the increasing temperature. The liquid flow is directed to the heater near the substrate, and it transfers the particles to the center of the cell. The volume fraction of the particles increases over time in this region. The heat flow from the heater affects the geometry of the cluster for two reasons: first, the Marangoni flow velocity depends on the temperature gradient, and second, the decrease in film thickness near the heater depends on the temperature. The results of the simulation are in general agreement with the experimental data.
\end{abstract}

\maketitle

\section{Introduction}

The method of evaporative lithography consists in creating
conditions for nonuniform evaporation from the free surface of a
colloidal liquid~\cite{RouthRussel1998,Harris2007,Kolegov2020}. The
compensatory flow occurs due to the nonuniform evaporation of the
liquid, and transfers the particles to the areas of intense
evaporation. It is possible to control the vapor flux density along
the free surface of the liquid layer, e.g., by placing a mask
above a droplet~\cite{Harris2007,Vodolazskaya2017}. The resulting
structures follow the shape of the holes in the masks, which play
the role of templates. In the case of a predominant Marangoni flow,
the precipitation structures are inverted relative to the holes in
the mask~\cite{HarrisLewis2008}. External sources of
airflow~\cite{Yang2021} and composite substrates with variable
thermal properties~\cite{Cavadini201324,Cavadini2015} can also be
useful for evaporative lithography applications. If the melting
point of the particles is higher than room temperature, the
resulting precipitation is brittle. Therefore, to obtain solid
relief films a heating step using  IR light is added to the
procedure~\cite{Routh2011,Georgiadis2013,Kolegov2018113}. In addition,
this step accelerates the drying process and the formation of a hard
coating. The effect of laser heating of the droplet apex on the
deposition structure of colloidal particles was compared with
the heating of the entire substrate, as reported in Ref~\cite{Yen2018}. A similar experiment was performed with a drop of saline solution on hydrophobic and hydrophilic substrates~\cite{Liu2020}. The process of droplet heating by xenon lamp radiation leads to a more pronounced annular deposit of gold nanoparticles, if compared with the heating of the substrate, when a part of the nanoparticles is deposited in the inner region~\cite{Yan2020}. Coffee rings formed during droplets drying on substrates at a temperature gradient are considered in the experiment~\cite{Malla2019}. It is possible to change the direction and pattern of flow due to the uniform or nonuniform heating of the droplet~\cite{Thokchom2015}. The experiment was carried out with a drop confined between two plates of glass (Hele--Shaw cell). The local particle concentration was measured at different points in time. The Marangoni flow was modeled using the stationary Navier--Stokes equations. The model takes into account thermodiffusion and heat convection. The temperature boundary
conditions were specified by linear approximations. The
experiment~\cite{Zaaroura2021} showed the effect of gold
nanoparticles on the evaporation rate of a drop on a heated
substrate. At different substrate temperatures, an annular
precipitate, a uniform spot~\cite{Lama2020}, or an eye-shaped
precipitate~\cite{Li2015} can form. This is due to the competition
between capillary and thermocapillary flows. It is then possible to use such competition to regulate the thickness of the micro-needles that
form when the polymer film dries in a heated cell, near its
wall~\cite{Mansoor2011}. The effect of the substrate temperature and
the initial concentration of nanoparticles on the final sediment
shape was previously studied~\cite{LiuB2020}. The geometrical
characteristics of drying hydrocarbon droplets and the resulting
deposits on a heated aluminum substrate were analyzed
in Ref~\cite{Haenichen2019}.

A number of papers have considered pure liquids. The theory of
Marangoni flows in nonuniformly heated films was described in detail
in the review~\cite{GambaryanRoisman2015}. The thermocapillary rupture of
the film in a cell with a built-in heater was experimentally investigated
in Refs~\cite{Spesivtsev2017,Kochkin2020,KochkinJPCS2020}. The kinetics of evaporation
under local heating of the droplet was studied in Ref~\cite{Askounis2017}.
Localized heating of the substrate with a laser was carried out
in the area of the center or periphery of the drop.
This has led to the appearance of Marangoni flows. Various thermal patterns associated with
hydrothermal waves have been previously observed in literature~\cite{Askounis2017}. Moreover, simulation results have shown that the evaporation is jointly affected by the local heating of the droplet and the wettability
of the substrate~\cite{Wang2020}. An experiment with the evaporation of a drop on a heated hydrophobic substrate has been carried out in Ref~\cite{Tam2009}. An analytical solution of the stationary problem for calculating the Marangoni flow has also been obtained. Droplet impact on a heated surface was studied experimentally in Ref~\cite{Gatapova2018}. The study of evaporation of droplets of the ethanol-water binary  mixture on a heated substrate has been carried out in Ref~\cite{Gurrala2019}. The influence of the topographic structure of the substrate and its temperature on the wetting mode of a solid surface with an oil drop during evaporation has been considered in Ref~\cite{Khilifi2020}.

In the present work, we study the possibility of controlling the particle assembly during the evaporation of the liquid they are dispersed in, using a local heater mounted in the central part of the bottom of the open cell. The peculiarity of this work is that the particles are relatively large and heavy, so they are not suspended in solution. Sedimentation prevails over diffusion, so at the beginning of the process, the particles are concentrated near the bottom of the cell. In our opinion, this way of forming particle clusters can be attributed to hybrid methods of evaporative lithography. According to the classification~\cite{Kolegov2020}, such methods combine the creation of conditions for nonuniform evaporation and other mechanisms related and not related to evaporative self-assembly. The temperature gradient of the free surface of the liquid can cause a capillary flow that compensates for the loss of liquid in areas of relatively rapid evaporation. The liquid turns into vapor faster where the temperature is higher. On the other hand, this temperature gradient can contribute to the appearance of thermocapillary convection, which is not associated with evaporation. This is the so-called Marangoni effect in which the convection of a liquid is associated with a surface tension gradient. So, in our conditions, the surface tension depends on the temperature of the liquid. The goal of the present work is to understand which of the two mechanisms prevails in the proposed experiment. Contrary to the procedure described in the previous work~\cite{AlMuzaiqer2021}, a higher solution concentration and relatively thick liquid layers are used here. This choice allows us to obtain multilayered clusters. Besides, in the present work, the dependence of the cluster height on the thickness of the liquid layer is studied, and the particle velocity field is measured.


\section{Methods}

\subsection{Experiment}

\subsubsection{Technical details}
\begin{figure}[hbt]
    \includegraphics[width=0.95\linewidth]{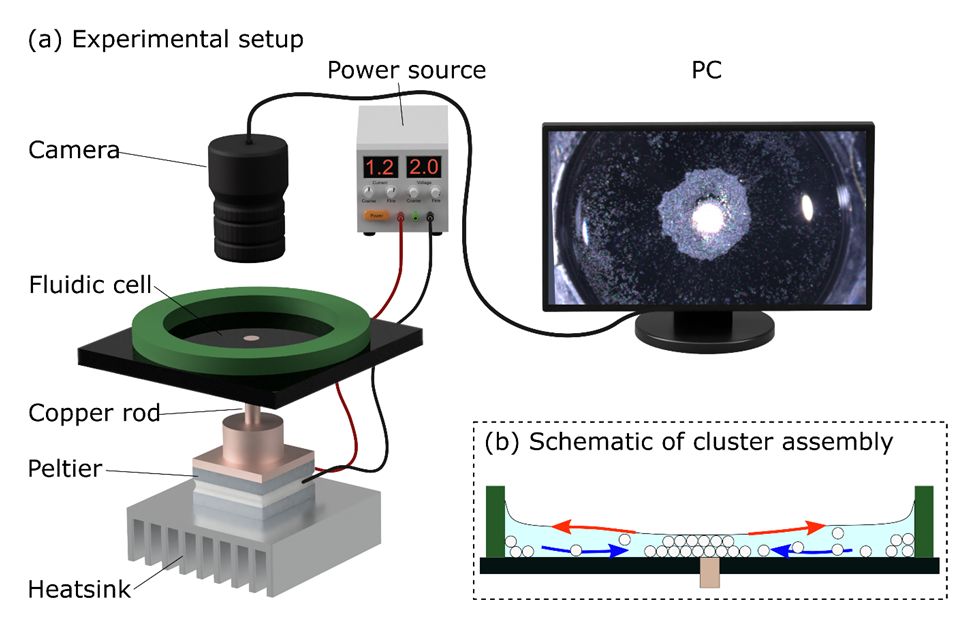}\\
    \caption{
        (a) Sketch of the experimental setup. (b) An illustration of the assembly process of the cluster in the heating zone. The outwards arrows show thermocapillary flows. The inwards arrows show reverse bottom flows towards the heater.
    }
    \label{fig:experimentSketch}
\end{figure}

A sketch of the experimental setup is shown in Fig.~\ref{fig:experimentSketch}a. The fluidic cell consists of a substrate of welding glass, with a photopolymer ring glued on top of it (ring dimensions: height = 4 mm, inner diameter = 20 mm). In the center of the substrate, a hole has been drilled. A copper rod (section radius of $R_h=$ 0.9~mm) was hermetically fitted into the hole and connected to the heated side of a Peltier element (NEC1-00703, 10$\times$10$\times$4.9 mm$^3$). As working materials, polystyrene microspheres and volatile isopropyl alcohol (isopropanol) were used. The geometrical parameters and properties of materials are summarized in Tables~\ref{table:GeometricParameters}, \ref{table:PhysicalParameters}.

\begin{table}[h]
\caption{Geometric parameters.}
\centering
\begin{tabular}{|l|l|l|}
  \hline
  Symbol & Parameter & Value \\
  \hline
  $d_p$ [$\mu$m] & Particle diameter & 50 \\
  $h_0$ [$\mu$m] & Initial film height & 700 \\
  $H_\mathrm{bot}$ [mm] & Cell bottom thickness & 3 \\
  $R_h$ [mm] & Heater radius & 0.6 \\
  $R_\mathrm{in}$ [cm] & Inner radius of the cell & 1 \\
  $\phi_0$ [--] & Initial volume fraction & 0.002 \\
  $\phi_\mathrm{max}$ [--] & Maximum volume fraction & 0.64 \\
  \hline
\end{tabular}
\label{table:GeometricParameters}
\end{table}

\begin{table}[ht]
\caption{Physical parameters.}
\centering
\begin{tabular}{|l|p{0.21\textwidth}|l|}
  \hline
  Symbol & Parameter & Value \\
  \hline
  $D$ [m$^2$/s] & Particle diffusion coefficient & $3.6\times 10^{-15}$ \\
  $c_l$ [J/(kg K)] & Specific heat capacity & 2605 \\
  $c_s$ [J/(kg K)] & Specific heat capacity & 780 \\
  $k_l$ [W/(m K)] & Thermal conductivity & 0.13 \\
  $k_s$ [W/(m K)] & Thermal conductivity & 0.748 \\
  $L$ [J/kg] & Heat of vaporization & $75\times 10^4$ \\
  $M$ [kg/mol] & Molar mass & 0.06 \\
  $P_v$ [Pa] & Partial pressure of the gas in the mixture & $4.23\times 10^3$ \\
  $q$ [W/m$^2$] & Heat flux density & $10^4$ \\
  $R$ [J/(kg K)] & Universal gas constant & 8.31 \\
  $T_0$ [K] & Room temperature & 300 \\
  $T_\mathrm{sat}$ [K] & Saturation temperature & 355.4 \\
  $\alpha_e$ [--] & Evaporation coefficient & $3.1\times 10^{-5}$ \\
  $\alpha_{ls}$ [W/(K m$^2$)] & Heat transfer coefficient & 69.3 \\
  $\eta_0$ [Pa s] & Viscosity & $2.43\times 10^{-3}$ \\
  $\rho_l$ [kg/m$^3$] & Liquid density & 786 \\
  $\rho_s$ [kg/m$^3$] & Substrate density & 2500 \\
  $\rho_p$ [kg/m$^3$] & Particle density & 1050 \\
  $\sigma'$ [N/(m K)] & $\sigma'=\partial \sigma/\partial T$ & $-8\times 10^{-5}$ \\
  $\sigma_0$ [N/m] & Surface tension & $22\times 10^{-3}$ \\
  \hline
\end{tabular}
\label{table:PhysicalParameters}
\end{table}

The experiments were carried out according to the following procedure: dry microparticles are placed onto the substrate of the experimental cell and covered with an isopropanol layer; the particles are stirred constantly with a pipette tip as long as particle distribution begins to be uniform in the layer; analysis of optical images of such initial distribution have shown that particles are quite densely cover the substrate; the particles are heavier than the liquid, so before the experiment begins, they lie at the bottom of the cell, that is, they have negative buoyancy; when the heating is started, the process of cluster formation is recorded, using a Axio Zoom.V16 microscope, with lens Zeissapoz 1.5x/0.37 (the free working distance 30 mm), equipped with CCD camera Zeiss Axiocam 506 color (Fig.~\ref{fig:experimentSketch}a). The camera is used in FullHD mode ($1920\times 1200$ pixels) at 30 frames per second (fps), and the spatial resolution obtained with the mentioned optics was about 17 $\mu$m/pixel. When the particles have formed the cluster (Fig.~\ref{fig:experimentSketch}b), isopropanol continues to evaporate  until it is completely exhausted. The cycle is repeated: a fixed  volume of alcohol is added into the cell, where the dry particles are. We have performed experiments using two liquid film thicknesses ($h_0\approx$ 400 and $h_0\approx$ 700~$\mu$m) at a constant voltage (+2V) applied to the Peltier module. To check the reproducibility, experiments have been repeated five times for each layer thickness. Then, for both values of $h_0$, the captured images have been analyzed  and the area of the particle cluster, $S(t)$, forming on the substrate upon heating has been measured. The measurement of the cluster profile, $\delta(r)$, has been performed using the contour analysis system  (the model OCA 15 DataPhysics Instruments). The evolution and radial distribution of temperature have been both measured using an IR camera (Flir A655sc, spectral range 7.5--14~$\mu$m, $\pm2^\circ$C) equipped  with a 0.3 Megapixel sensor ($640\times 480$ pixels), at 25 fps. The spatial resolution obtained with a 80~mm objective has been 38.4 $\mu$m/pixel.

\subsubsection{Cluster area estimation}

\begin{figure}[hbt]
    \includegraphics[width=0.95\linewidth]{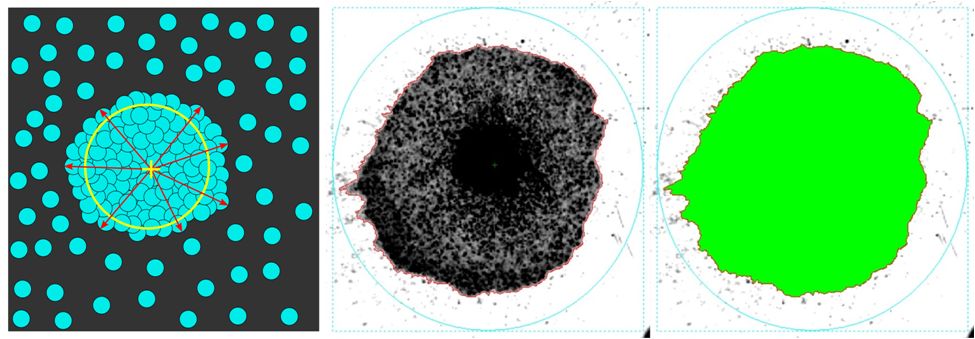}\\
    \begin{minipage}{0.3\linewidth}
       \center (a)
     \end{minipage}
     \begin{minipage}{0.3\linewidth}
        \center (b)
     \end{minipage}
     \begin{minipage}{0.3\linewidth}
        \center (c)
     \end{minipage}
    \caption{
     Method for measurement of the cluster area: (a) schematic representation of the cluster, (b) formation of the cluster outline (red line), (c) calculation of the cluster area (green area enclosed by the red line).
    }
    \label{fig:inverseVision}
\end{figure}

During the heating, the evolution of the area of the particle cluster was measured, using a sequence of frames. To this end, the boundary of the cluster has been defined based on a transition from high to low intensity of pixels in each digital image. A schematic of the method for determining the boundary is shown in Fig.~\ref{fig:inverseVision}a. The colors of images are inverted to improve the perception. A point, marked  by a yellow cross, is set in the center of the growing cluster. Then, lines are drawn from this point (see red arrows in Fig.~\ref{fig:inverseVision}a) toward all directions with an angular step of 0.25$^\circ$. Such a small angle makes it possible to measure in detail the boundary of a cluster in an image with a dimension of $1920 \times 1200$ pixels.
The length of the line is determined by the outermost particle of the cluster. We have used the linear interpolation of the outer cluster boundary based on the obtained discrete values (the red outline in Fig.~\ref{fig:inverseVision}b). Then, the area (the green area in Fig.~\ref{fig:inverseVision}c) enclosed by that boundary is calculated.

\subsubsection{Error analysis}

The error in determining the cluster area is the sum of the error in determining the pixels included in the relevant area plus the statistical error arising from the repeating and averaging of a series of experiments with fixed parameters. The first type of error in measuring the area is associated with 1) inaccuracy in determining the cluster boundary and 2) overestimation of the cluster area due to particles moving towards the center.
It was found that individual particles form in the images a spot with a diameter of 4$\pm$1 pixels (the image resolution is 17~$\mu$m~/ pixel), which means that the error in determining the position of the cluster boundary can be estimated as being $\pm$0.5 pixel. To estimate the error in measuring the area, the cluster perimeter has been measured at three points in time after the start of heating: 20, 30, and 45 s. Later, we have analyzed other error components considering these time periods.  The perimeter has been calculated as the number of pixels on the cluster border. Therefore, taking into account the error in determining the position of the border, the error in the area measurement is equal to half the area of pixels located along the perimeter. The estimates for the considered points in time have shown that the maximum value of the relative error in the area measurement is $\pm$1.4\%.

An overestimation of the cluster area arises if the particle transferred by the flow over the stationary particles crosses the cluster boundary. In the current frame, the area of such particle is added to the total cluster area, but in the next frames, they can move above the cluster and, as a result, not increase the measured cluster area.
This error contributes at the early stages noticeably due to the small size of the forming cluster and at the later stages due to the high intensity of convective flows. It is possible to estimate the largest value of this error by multiplying the perimeter of the cluster by the particle radius. This value corresponds to the case where the cluster is surrounded by such particles, but some of them are just approaching the cluster boundary, and others are crossing the cluster boundary. In reality, there is a gap between moving particles, which is not smaller  than the particle diameter. As a result of this error estimate, we have corrected our computation by dividing the perimeter in half. The maximum value of the relative error is $\pm$5.6\% in this case. The final values of the error in measuring the area do not exceed $\pm$5.8\%.

\subsubsection{Particle velocity field}

The cluster assembly process is realized by the movement of microparticles toward the heater. The video recording of this movement can be analyzed using particle image velocimetry (PIV) technology, which allows to determine the particle velocity field. However, in contrast to the standard PIV measurements that assume a small tracer concentration, in our experiments there is a huge particle number. This increases as a consequence the probability of wrong determination of the particle positions and their velocities. In addition, individual particles are detected only outside the cluster, while particles moving above it are not detected against the background of nonmoving particles in the cluster. It should also be noted that with the development of a thermocapillary vortex and an increase in the flow velocity, the number of particles entrained by the outward thermocapillary flow along the free surface towards the wall of the cell increases. Therefore, the map of particle velocity field contains vectors directed both to and from the heater, which complicates the processing. For this reason, it is possible to obtain plausible results only for the initial stage of the particle assembly process (during 60~s from the beginning of heating).

To determine the velocity field, the video was split into frames with a frequency of 10 Hz and analyzed using the open-source software package (OpenPIV)~\cite{BenGida2020}.
Then, the velocity fields obtained at each point in time were processed by averaging the velocity magnitude over the radius relative to the center of the heater, with a step of 0.1~mm. We saved the obtained spatiotemporal values of the flow velocity in a two-dimensional array for further visualization.

\subsection{Simulation}\label{simulation}
\subsubsection{Problem statement}\label{subsubsec:problemDefinition}
\begin{figure}[hbt]
    \includegraphics[width=0.8\linewidth]{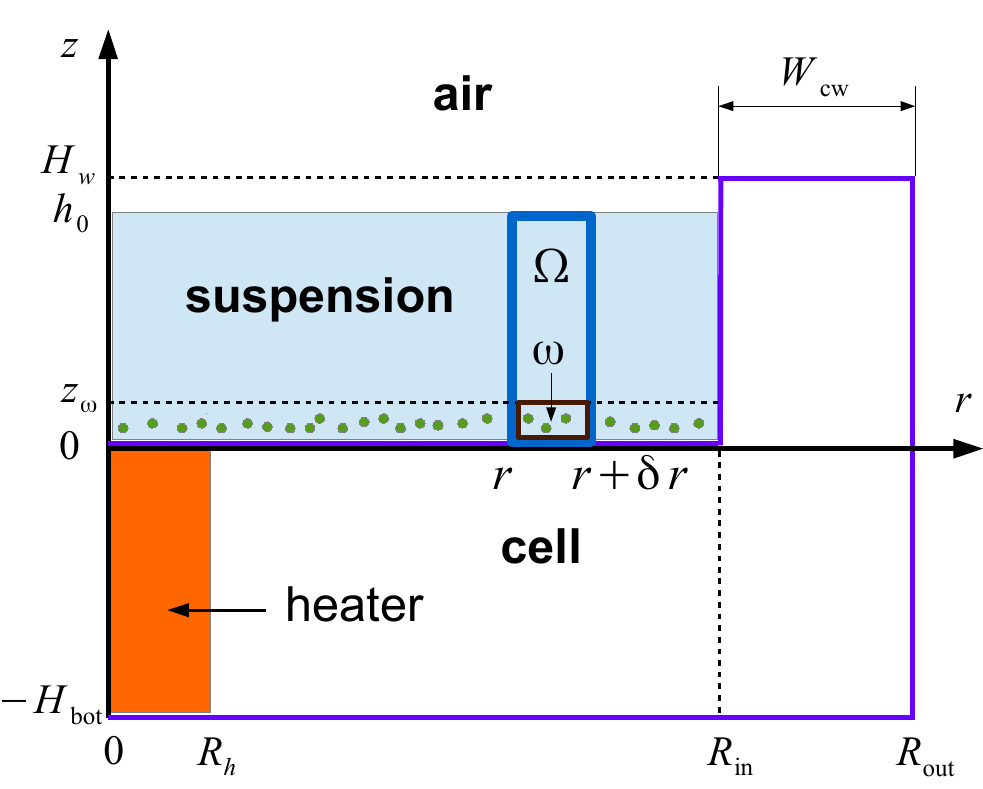}\\
    \caption{The domain of the problem solution.}
    \label{fig:domain}
\end{figure}

Let us consider a suspension in an open glass cell
(Fig.~\ref{fig:domain}). A heater is mounted in the center of the bottom of the cylindrical cell. We denote the inner radius of the cell as $R_\mathrm{in}$, the radius of the mounted heating element as $R_h$, and the initial thickness of the film as $h_0$. The film is thin ($h_0 \ll R_\mathrm{in}$), so the vertical transfer of mass and
heat is not taken into account. In addition, as the experiment has shown, the particles move mainly near the substrate.

Here we focus on a qualitative description of the process: we
believe that the particle density corresponds to the density of the liquid.  In
fact, the density of isopropanol is about 20\% less than the
density of polystyrene.

\subsubsection{Governing equations}\label{subsubsec:GoverningEquations}
The fluid flow velocity will be calculated using the lubrication approximation~\cite{Sultan2005,Yamamura2009,Cazabat2010},
\begin{equation}\label{eq:velocity}
  u= \frac{h}{2\eta_0}\frac{\partial \sigma}{\partial r}-\frac{h^2}{3\eta_0}\frac{\partial P}{\partial r},
\end{equation}
where $h(r,t)$ is the film thickness, $u(r,t)$ is the velocity of the radial flow averaged by the film thickness, $\eta_0$ is the dynamic viscosity of a pure liquid, $\sigma(r,t)$ is the surface tension coefficient, $P(r,t)$ is the capillary pressure, and $r$ and
$t$ are the space coordinate and time accordingly.  By using the linear approximation  $\sigma=\sigma_0 + \sigma'
\Delta T$, we get $\partial \sigma/\partial r=\sigma'
\partial T/\partial r$, where $\Delta T = T_l - T_0$, $T_l(r,t)$ is the liquid temperature,
$\sigma_0$ is the surface tension at room temperature $T_0$, and $\sigma'=\partial \sigma/\partial T$. The capillary pressure $P(r,t)$ is described by the Young--Laplace equation. Taking into account the lubrication approximation~\cite{Fischer2002}, we have
\begin{equation}\label{eq:CapillaryPressure}
P=-\frac{\sigma}{r}\frac{\partial}{\partial r}\left( r
\frac{\partial h}{\partial r} \right).
\end{equation}
Equation~\eqref{eq:velocity} takes into account the capillary flow and the Marangoni flow. The evolution in the film thickness in space and time is described using the law of mass conservation
\begin{equation}\label{eq:conservationMassLaw}
 \frac{\partial h}{\partial t} +\frac {1}{r} \frac{\partial}{\partial r}\left( rhu (1
-\phi) \right) + \frac{1}{r} \frac{\partial}{\partial r}(r h
U_\omega \phi) =-\frac{J}{\rho_l},
\end{equation}
where $\phi(r,t)$ is the particle volume fraction averaged by the film thickness, $\rho_l$ is the liquid density, and $U_\omega$ is the fluid flow velocity in subdomain $\omega$. The vapor flux density $J(r,t)$ is determined using the  Hertz--Knudsen formula~\cite{Davis1975,Semenov2013,Persad2016,Bekezhanova2018,Zigelman2019,Zhang2019}
$$J=\alpha_e \sqrt{\frac{M}{2\pi R T_\mathrm{sat}}} \left( P_\mathrm{sat}(T)-P_v \right).$$
Here, $\alpha_e$ is the evaporation coefficient~\cite{Gerasimov2018}, $M$ is the molar mass,
$R$ is the universal gas constant, $T_\mathrm{sat}$ is the
saturation temperature, $P_\mathrm{sat}$ is the saturated vapor pressure, and $P_v$ is the partial pressure of the gas in a mixture. Let us write the flow velocity in the subdomain $\omega$ as~\cite{Marchuk2009,Gaalen2021}:
\begin{equation}\label{eq:velocityInOmega}
  U_\omega= \frac{z_\omega}{\eta_\omega}\frac{\partial \sigma}{\partial r}+\frac{(z_\omega)^2}{2\eta_\omega}\frac{\partial P}{\partial r}-
  \frac{z_\omega h}{\eta_\omega}\frac{\partial P}{\partial r},
\end{equation}
where $\eta_\omega(r,t)$ is the dynamic liquid viscosity. The dependence of the viscosity on the solution concentration is predicted by Mooney's formula~\cite{Mooney1951}:
$$ \eta_\omega= \eta_0 \exp \left( \frac{2.5 \phi_\omega}{1-K \phi_\omega} \right),$$
where $\phi_\omega(r,t)$ is the volume fraction of particles in the subdomain
$\omega$ and parameter $K=1/\phi_\mathrm{max}$ ($\phi_\mathrm{max}$ is the
maximum volume fraction of particles). In the following calculations, we have used the value $z_\omega= d_p$, where $d_p$ is the diameter of the particles.

Particle transfer is described by the convection--diffusion equation:
\begin{equation}\label{eq:convectionDiffusion}
  \frac{\partial (\phi h)}{\partial t} + \frac{1}{r} \frac{\partial (r U_\omega h \phi)}{\partial r}
  =  \frac{D}{r} \frac{\partial}{\partial r}\left( r \frac{\partial (\phi h)}{\partial r} \right).
\end{equation}
The equation
\begin{multline}\label{eq:HeatTransferInLiquid}
 \frac{\partial (hT_l)}{\partial t} + \frac{1}{r} \frac{\partial
(rhuT_l)}{\partial r}= \\ \frac{k_l}{c_l \rho_l} \frac{1}{r}
\frac{\partial}{\partial r}\left( rh \frac{\partial T_l}{\partial r}
\right) +  \alpha_{ls} \frac{T_s - T_l}{\rho_l c_l} -
\frac{LJ}{\rho_l c_l}-\frac{J T_l}{\rho_l}
\end{multline}
describes heat transfer in liquid, where
$k_l$ is the thermal conductivity of the liquid, $c_l$ is the
specific heat capacity, $L$ is the heat of vaporization,
$\alpha_{ls}$ is the convective heat transfer coefficient between
liquid and substrate, and $T_s(r,t)$ and $T_l(r,t)$ are substrate and liquid temperatures averaged by the thickness of substrate and film, respectively. Heat transfer equation~\eqref{eq:HeatTransferInLiquid} takes into account the convective heat transfer in the liquid, the thermal conductivity of the liquid, the heat exchange with the substrate, and the cooling due to evaporation. The evolution of the substrate temperature is described by the thermal conductivity equation
\begin{equation}\label{eq:HeatTransferInSolid}
  \frac{\partial T_s}{\partial t} =
  \frac{k_s}{\rho_s\, c_s} \frac{1}{r}
  \frac{\partial}{\partial r}\left( r \frac{\partial T_s}{\partial r} \right)
  + \frac{\alpha_{ls}}{\rho_s\, c_s} \frac{T_l - T_s}{H_\mathrm{bot}}+\frac{q}{\rho_s\, c_s\, H_\mathrm{bot}},
\end{equation}
where the heat flux density function of $q=q_0 f_\mathrm{sh}(r)$.
Here, $q_0$ is the heat flux density from the heater and we have used the smoothing function of
 $f_\mathrm{sh}(r)=-p_s\, r^2 / R_\mathrm{in}^2$, where $p_s= R_\mathrm{in}/R_h$.
Equation~\eqref{eq:HeatTransferInSolid} takes into account the thermal conductivity, the heat source, and the heat exchange between the substrate and the liquid.

The derivation of equations~\eqref{eq:conservationMassLaw},
\eqref{eq:convectionDiffusion},
\eqref{eq:HeatTransferInLiquid}, and  \eqref{eq:HeatTransferInSolid} is given in the \hyperref[sec:appendix]{appendix}.

\subsubsection{Initial and boundary conditions}\label{InitialBoundaryConditions}
Let us write down the initial conditions of the problem
\begin{equation}\label{eq:InitialConditions}
  h(r,0)=h_0, T_{l,s}(r,0)=T_0, \phi(r,0)=\phi_0,
\end{equation}
where $T_0$ and $\phi_0$ are the specified constants.
In~\eqref{eq:InitialConditions}, we have considered a flat film surface, a uniform distribution of the particle concentration and the liquid temperature at time $t=0$. In fact, the film is similar to the meniscus, as the wall of the cell is wetted with a liquid. This shape of the free surface is not taken into account here since a significant curvature of the two-phase boundary is observed far away from the heating element at $r>0.8R_\mathrm{in}$, while the cluster is formed in the central part of the cell.

The boundary conditions are
\begin{multline}\label{eq:BoundaryConditions}
  \frac{\partial h(R_\mathrm{in},t)}{\partial r}=0,
  \frac{\partial \phi(R_\mathrm{in},t)}{\partial r}=0,\\ \frac{\partial T_{l,s}(0,t)}{\partial r}=0,
  \frac{\partial T_{l,s}(R_\mathrm{in},t)}{\partial r}=0,\\
  u(0,t)=u(R_\mathrm{in},t)=U_\omega (0,t)=U_\omega (R_\mathrm{in},t)=0.
\end{multline}
The conditions in~\eqref{eq:BoundaryConditions} are written from the following considerations. We take into account that the radial component of the velocity vector tends to zero near the axis of symmetry and near the wall of the cell. The other conditions at $r=0$ are written taking into account the axial symmetry. The transfer of particles through the wall of the cell does not occur. In addition, we do not take into account the heat exchange with the environment at the boundary $r=R_\mathrm{in}$, since $R_\mathrm{in}\gg R_h$.  Here we consider only a special case of wetting of the wall with liquid, with a contact angle of $\theta = \pi/2$. Note that the boundary condition  $\partial
h(R_\mathrm{in},t)/\partial r=\cot \theta=0$ does not contradict the initial condition $h(r,0)=h_0$.

\subsubsection{Problem parameters}\label{subsubsec:parameters}

The geometric parameters of the problem are shown in the Table~\ref{table:GeometricParameters}. The value $\phi_\mathrm{max}$ corresponds to the volume fraction for a random dense packing of spheres. At first, the experiment was carried out for rods with $R_h\approx$ 0.6~mm. Then, rods with $R_h\approx$ 0.9~mm have been  used due to technical difficulties with drilling a hole in the substrate. But, for the calculations, we left the previous value of $R_h$ since no qualitative differences have been observed.

Table~\ref{table:PhysicalParameters} describes the physical parameters: indices $l$ and $s$ correspond to a liquid and a solid, respectively. The value $\alpha_e$ has been experimentally measured in this work and has been measured to be 30 times less than the value for a drop of isopropanol~\cite{Murisic2011}. The particle diffusion coefficient is calculated using the Einstein formula $D=k_B T_0/ (6\pi \eta_0 r_p)$, where
Boltzmann constant $k_B=1.38\times 10^{-23}$ J/K and the particle radius of
$r_p = 0.5 d_p$. The value of the heat flux density $q$ is selected approximately so that the rate of temperature growth in the central region approximately corresponds to the experimental data. The heat transfer coefficient is calculated approximately by the formula $\alpha_{ls}\approx k_l/ (0.5 H_\mathrm{bot} + 0.5 h_0)$. The rest of the information is taken from various reference books. The saturated vapor pressure has been calculated using the empirical formula $$P_\mathrm{sat}=A_2 + \frac{A_1 - A_2}{1+ \exp \left( (T_l - x_0)/ \Delta_x \right)},$$ where the parameters of $A_1=-1762.65199552881$, $A_2=503914.770838647$, $x_0=383.243463104184$, and $\Delta_x=20.494618894658$ have been used. These parameter values have been  determined using the tabulated data of $P_\mathrm{sat}$ and the least squares method. The calculations have used parameter values from the tables~\ref{table:GeometricParameters} and \ref{table:PhysicalParameters} by default, unless other values are specified in the text or in the plots.

\subsubsection{Numerical method}\label{subsubsec:NumericalMethod}
The system of equations~\eqref{eq:velocity}--\eqref{eq:HeatTransferInSolid} with initial and boundary conditions~\eqref{eq:InitialConditions},
\eqref{eq:BoundaryConditions} has been solved using the finite difference method.
The space derivatives were approximated with the central differences. We have used a constant step in time, $\Delta t=0.5$~s, and space, $\delta r = 10^{-4}$~m. The mesh convergence test has been  conducted. The implicit difference scheme of the first- and second-order approximation in time and space, respectively, has been solved using the Newton method. The work has been done using the  mathematical package Maple 2019. Unfortunately, the numerical implementation has allowed us to simulate the process only at the initial stage. We have been able to perform the calculation up to the time $t=25$~s. For greater values of time we have observed a numerical instability. Most likely, this is due to a small value of the diffusion coefficient $D$ and, as a consequence, a large concentration gradient $\phi$. A similar difficulty has occurred before in another work, where a different numerical method and a different software have been used~\cite{VieyraSalas2012}.

\section{Results and discussion}
\label{results}

\subsection{Experimental results}

\begin{figure}[hbt]
    \begin{minipage}{0.49\linewidth}
       \includegraphics[width=0.99\linewidth]{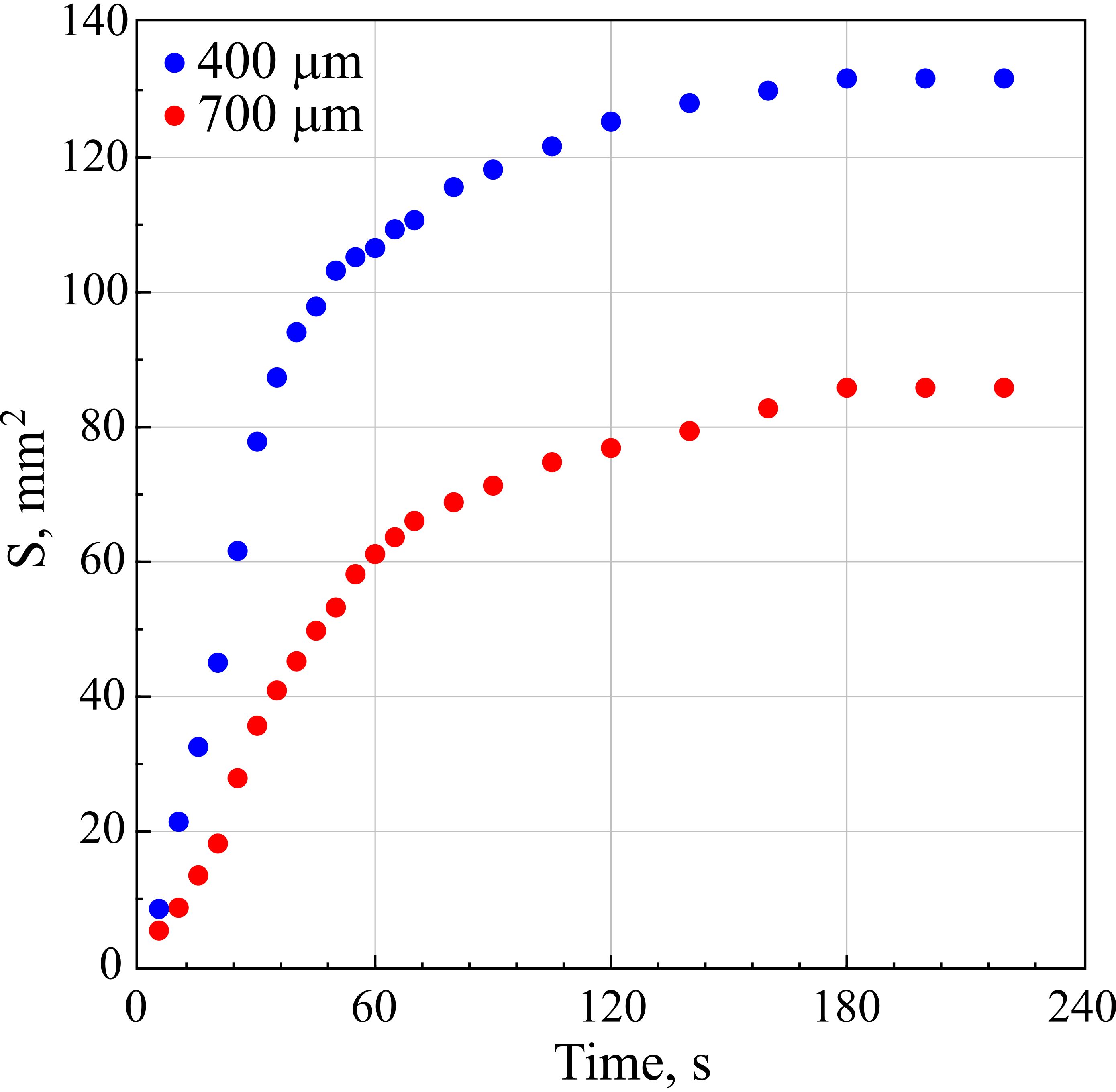}\\
       \center (a)
     \end{minipage}
     \begin{minipage}{0.49\linewidth}
       \includegraphics[width=0.99\linewidth]{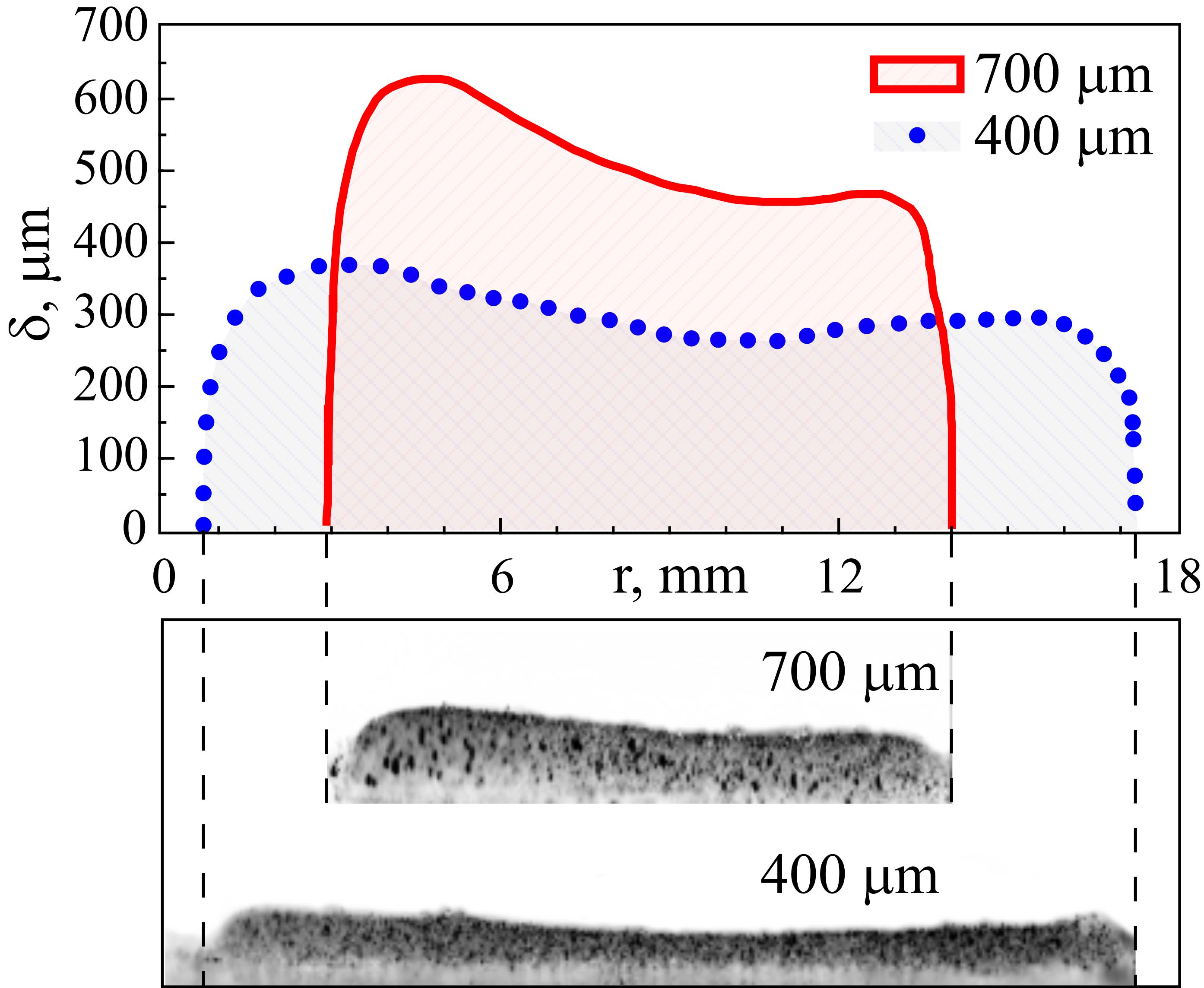}\\
        \center (b)
     \end{minipage}
    \caption{
        (a) Dependency of the cluster area on heating time at the layer thicknesses of 400 and 700~$\mu$m. (b) Top: the height of the cluster versus its diameter (the real scale). Bottom: side view images of the cluster for both layer thicknesses; the vertical scale of images is increased two times relative to the initial images.
    }
    \label{fig:SquarePatternProfile}
\end{figure}

Figure~\ref{fig:SquarePatternProfile}a shows the evolution of the cluster area in the layers of 400 and 700~$\mu$m thickness. The cluster area is smaller in the case of a thick layer compared to the case of a thin layer, which agrees with the trend found in our previous study for a small particle number~\cite{AlMuzaiqer2021}. Figure~\ref{fig:SquarePatternProfile}b shows the radial changing of the cluster height for both layers. The surface profiles were obtained by visual  contouring in side-view images of the cluster. The cluster is higher in the case of the 700~$\mu$m layer. Consequently, multilayer assemblies are formed in relatively thick layers. The reason for that is the particle uprising onto the cluster, caused by the action of upward flows at the liquid-cluster boundary~\cite{AlMuzaiqer2021}. It is notable that the time needed for the cluster to be formed, i.e., the assembly time of all particles in the heating spot is the same for both thicknesses of the layer.

\begin{figure}[hbt]
       \includegraphics[width=0.9\linewidth]{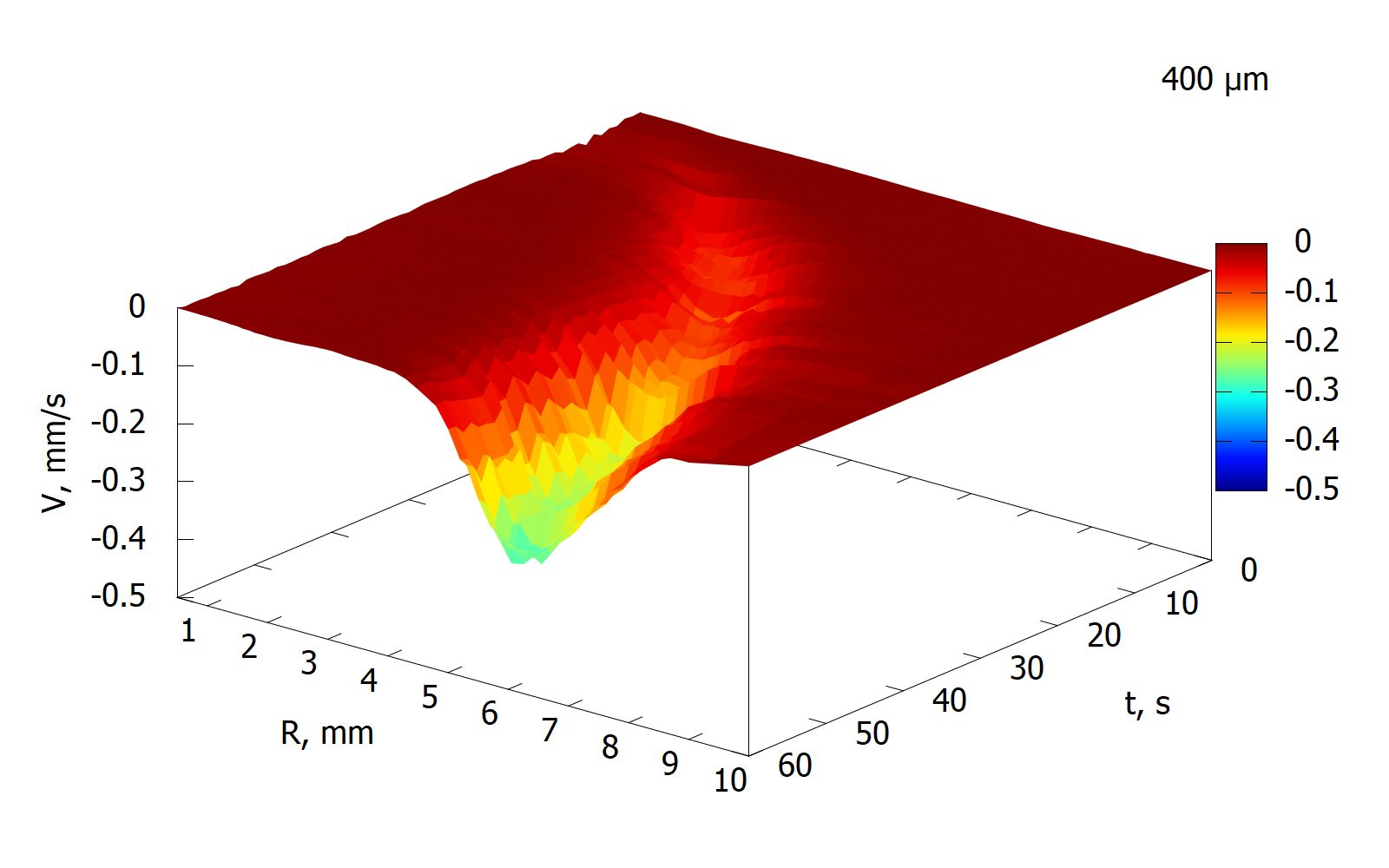} (a)\\
       \includegraphics[width=0.9\linewidth]{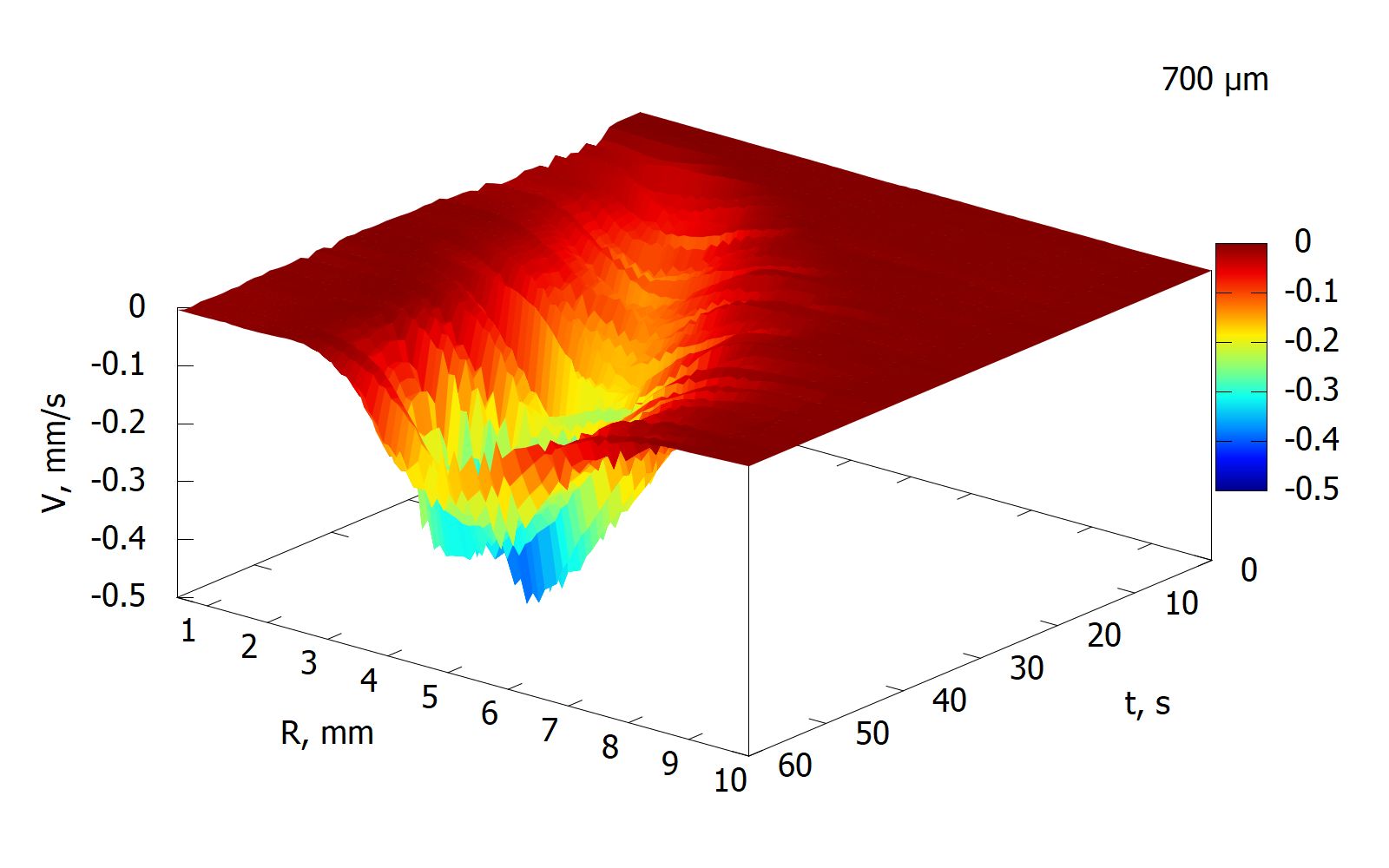}(b)
    \caption{Particle velocity in layers of 400~$\mu$m (a) and 700~$\mu$m (b).}
    \label{fig:particleVelocity}
\end{figure}

Figure~\ref{fig:particleVelocity} shows the velocity of particles entrained by the reverse bottom flow of liquid toward the heater.
The direction from the heater center towards the wall of the cell is chosen for the positive direction of the axis $r$. The maximum values of the velocity are achieved near the boundary of the growing cluster. This is obviously associated with an increase in the temperature gradient near the boundary compared to the periphery and, as a consequence, an increase in the flow velocity.

\begin{figure}
    \begin{minipage}{0.49\linewidth}
       \includegraphics[width=0.99\linewidth]{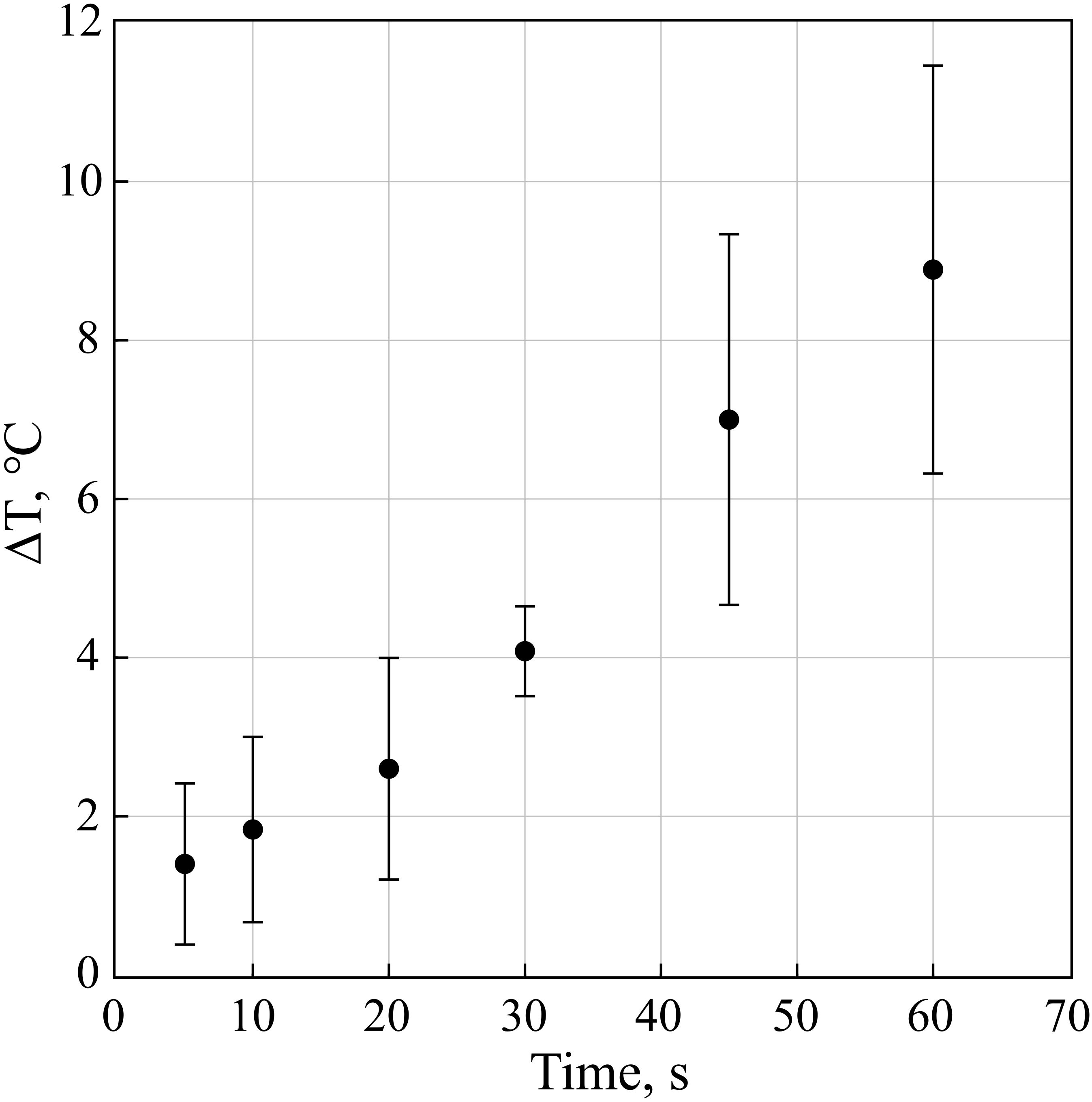}
       (a)\\
     \end{minipage}
     \begin{minipage}{0.49\linewidth}
       \includegraphics[width=0.99\linewidth]{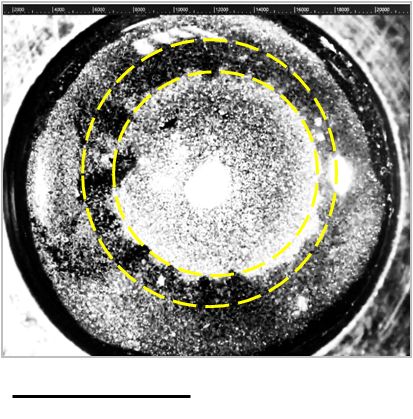}
        (b)
     \end{minipage}
    \caption{
    (a) Evolution of the temperature difference between the cluster edge and the wall. (b) A snapshot of the cluster assembly process ($t=$ 24~s) in the layer of 400~$\mu$m. The area bounded by the two rings is marked with dashed lines. The scale bar is 5~mm (Multimedia view).
    }
    \label{fig:TemperatureDifferenceEmptyRing}
\end{figure}

Interestingly also that the value of the maximum particle velocity increases with the heating time and the cluster area (Fig.~\ref{fig:particleVelocity}). This effect is caused by two mechanisms: an increase in the temperature difference between the cluster edge and the wall (Fig.~\ref{fig:TemperatureDifferenceEmptyRing}a), which leads to an increase in velocity, and the formation of a small ring-like zone near the cluster boundary, which eventually becomes relatively free of particles (Fig.~\ref{fig:TemperatureDifferenceEmptyRing}b, multimedia view), allowing particles to move freely in the flow without collisions or deceleration. The broadening of the averaged velocity profile over time in the 700~$\mu$m layer (Fig.~\ref{fig:SquarePatternProfile}b) is explained by the fact that the cluster boundary in that case is asymmetric in shape compared to the 400~$\mu$m layer.

\subsection{Simulation results}
\begin{figure}[hbt]
    \includegraphics[width=0.65\linewidth]{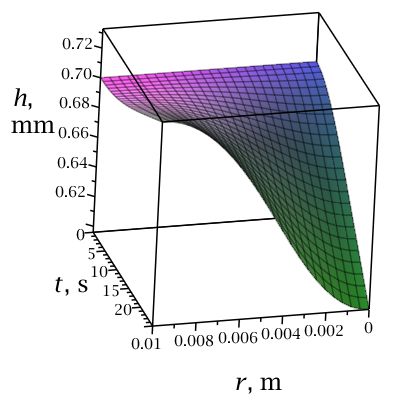}(a)\\
    \includegraphics[width=0.65\linewidth]{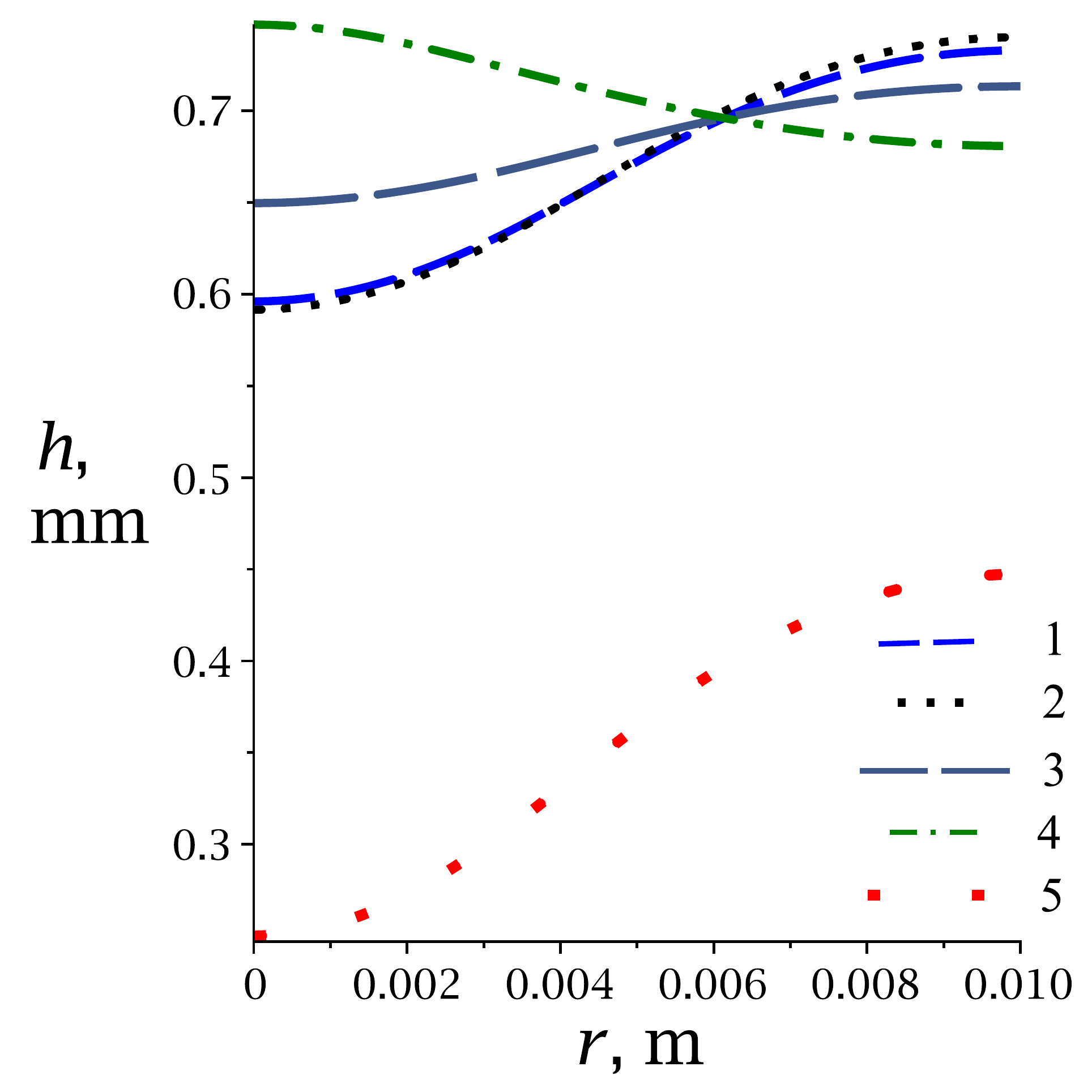}(b)
    \caption{Simulation results of (a) the spatiotemporal evolution of the film thickness and (b) the shape of the free liquid surface at time $t=25$~s for different values of the parameters: 1) $h_0=$ 700~$\mu$m, $q=$ 10$^4$~W/m$^2$; 2) $h_0=$ 700~$\mu$m, $q=$ 10$^4$~W/m$^2$, $\alpha_e=0$; 3) $h_0=$ 700~$\mu$m, $q=$ $0.5 \times 10^4$~W/m$^2$; 4) $h_0=$ 700~$\mu$m, $q=$ $-0.5 \times 10^4$~W/m$^2$; 5) $h_0=$ 400~$\mu$m, $q=$ $0.5 \times 10^4$~W/m$^2$.}
    \label{fig:heater_h}
\end{figure}

\begin{figure}[hbt]
    \includegraphics[width=0.65\linewidth]{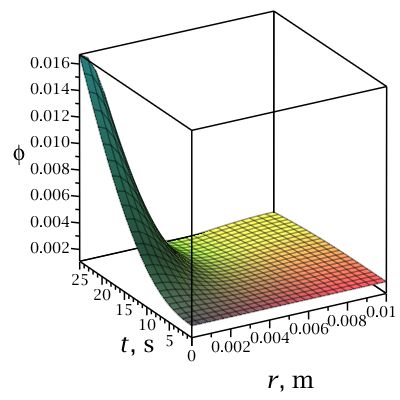}(a)\\
    \includegraphics[width=0.65\linewidth]{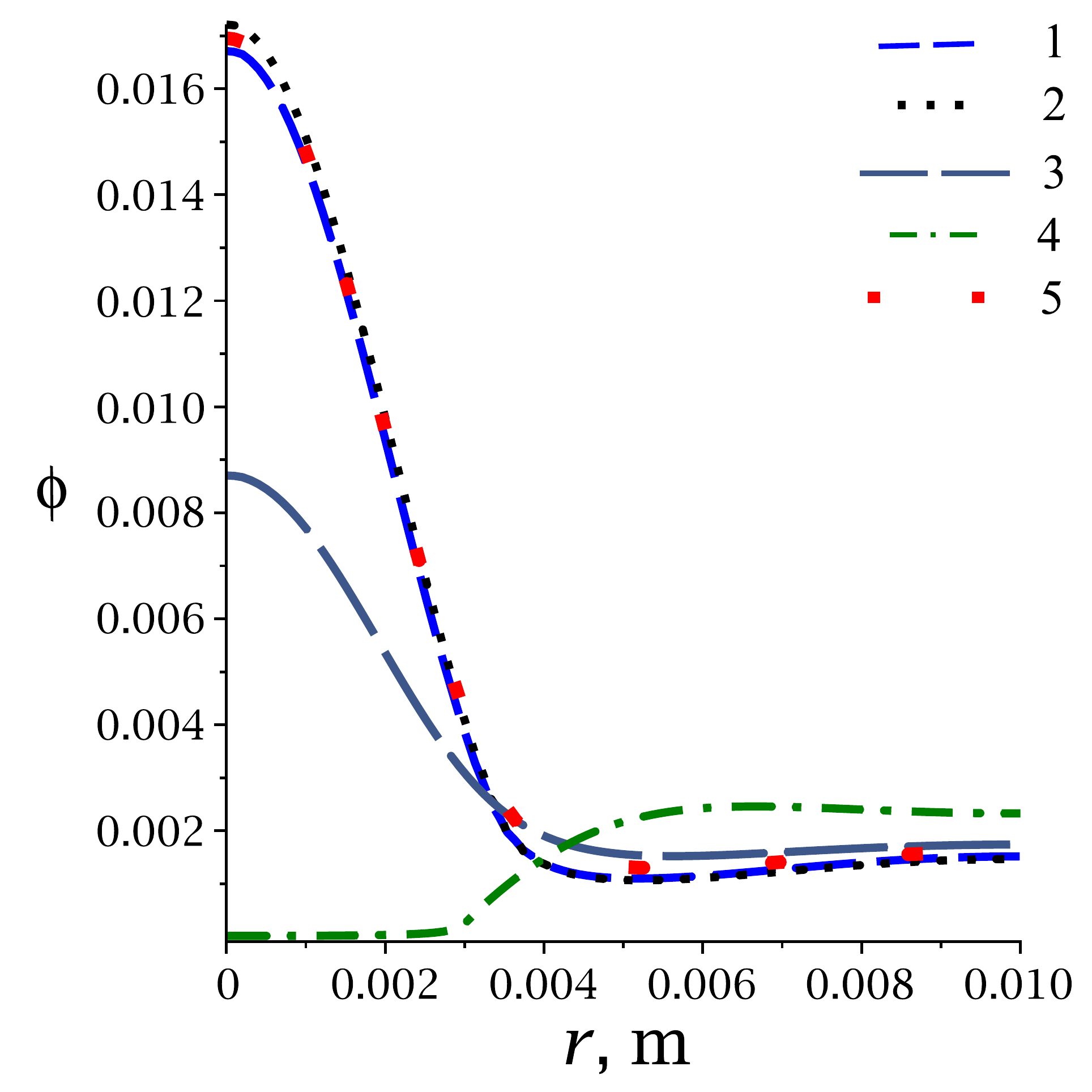}(b)
    \caption{Simulation results of (a) the spatiotemporal evolution of the volume particle fraction and (b) the spatial dependence of $\phi$ at time $t=25$~s for different values of the parameters: 1) $h_0=$ 700~$\mu$m, $q=$ 10$^4$~W/m$^2$; 2) $h_0=$ 700~$\mu$m, $q=$ 10$^4$~W/m$^2$, $\alpha_e=0$; 3) $h_0=$ 700~$\mu$m, $q=$ $0.5 \times 10^4$~W/m$^2$; 4) $h_0=$ 700~$\mu$m, $q=$ $-0.5 \times 10^4$~W/m$^2$; 5) $h_0=$ 400~$\mu$m, $q=$ $0.5 \times 10^4$~W/m$^2$.}
    \label{fig:heater_phi}
\end{figure}

\begin{figure}[hbt]
    \includegraphics[width=0.65\linewidth]{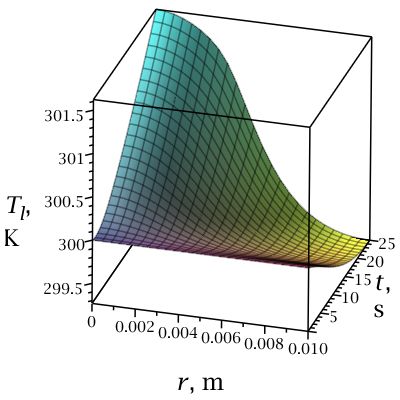}(a)\\
    \includegraphics[width=0.65\linewidth]{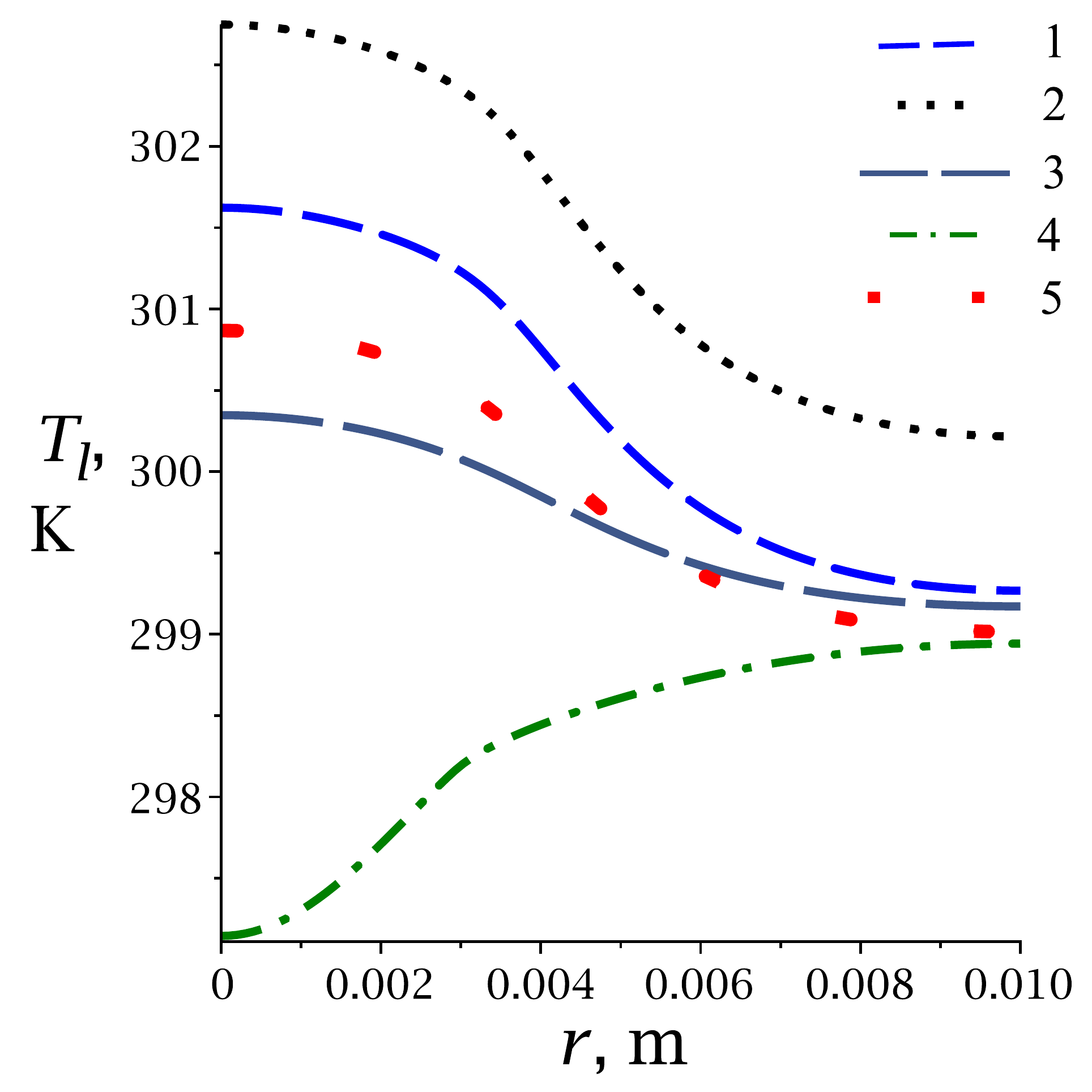}(b)
    \caption{Simulation results of (a) the spatiotemporal evolution of the liquid temperature and (b) the spatial dependence of $T_l$ at time $t=25$~s for different values of the parameters: 1) $h_0=$ 700~$\mu$m, $q=$ 10$^4$~W/m$^2$; 2) $h_0=$ 700~$\mu$m, $q=$ 10$^4$~W/m$^2$, $\alpha_e=0$; 3) $h_0=$ 700~$\mu$m, $q=$ $0.5 \times 10^4$~W/m$^2$; 4) $h_0=$ 700~$\mu$m, $q=$ $-0.5 \times 10^4$~W/m$^2$; 5) $h_0=$ 400~$\mu$m, $q=$ $0.5 \times 10^4$~W/m$^2$.}
    \label{fig:heater_Tl}
\end{figure}

\begin{figure}[hbt]
    \includegraphics[width=0.65\linewidth]{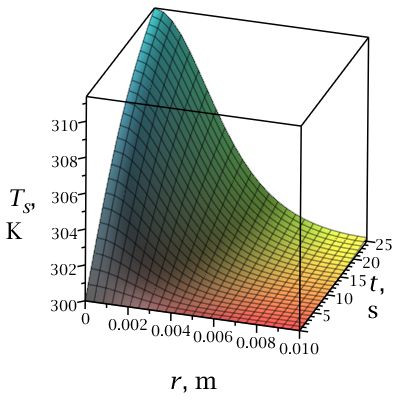}(a)\\
    \includegraphics[width=0.65\linewidth]{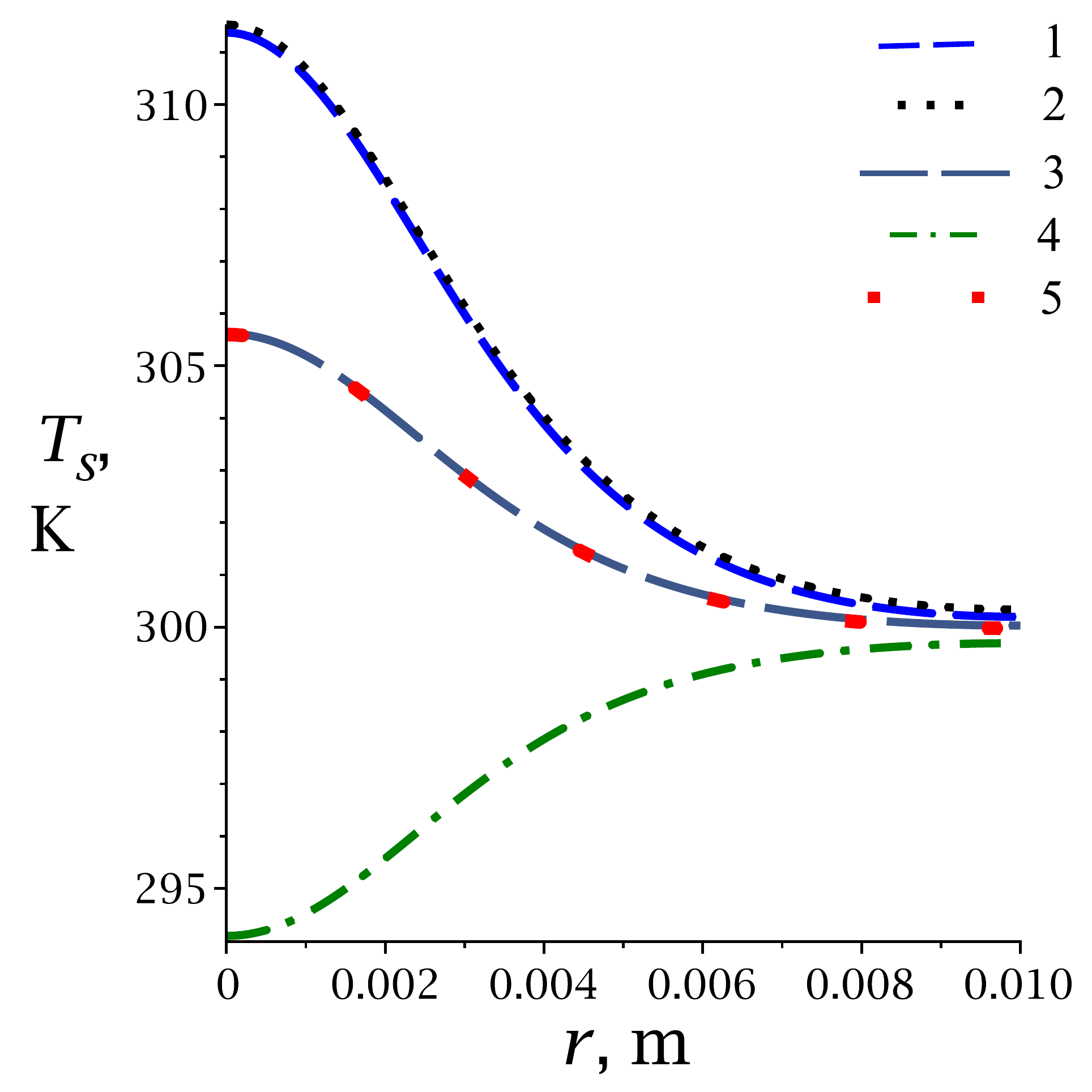}(b)
    \caption{Simulation results of (a) the spatiotemporal evolution of the substrate temperature and (b) the spatial dependence of $T_s$ at time $t=25$~s for different values of the parameters: 1) $h_0=$ 700~$\mu$m, $q=$ 10$^4$~W/m$^2$; 2) $h_0=$ 700~$\mu$m, $q=$ 10$^4$~W/m$^2$, $\alpha_e=0$; 3) $h_0=$ 700~$\mu$m, $q=$ $0.5 \times 10^4$~W/m$^2$; 4) $h_0=$ 700~$\mu$m, $q=$ $-0.5 \times 10^4$~W/m$^2$; 5) $h_0=$ 400~$\mu$m, $q=$ $0.5 \times 10^4$~W/m$^2$.}
    \label{fig:heater_Ts}
\end{figure}

\begin{figure}[hbt]
    \includegraphics[width=0.65\linewidth]{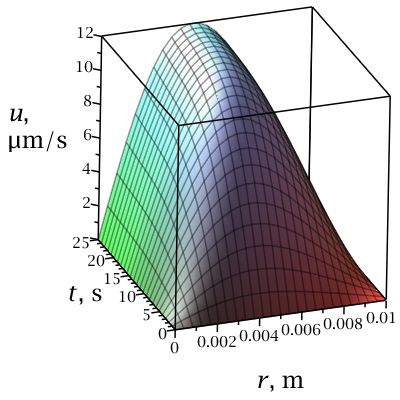}(a)\\
    \includegraphics[width=0.65\linewidth]{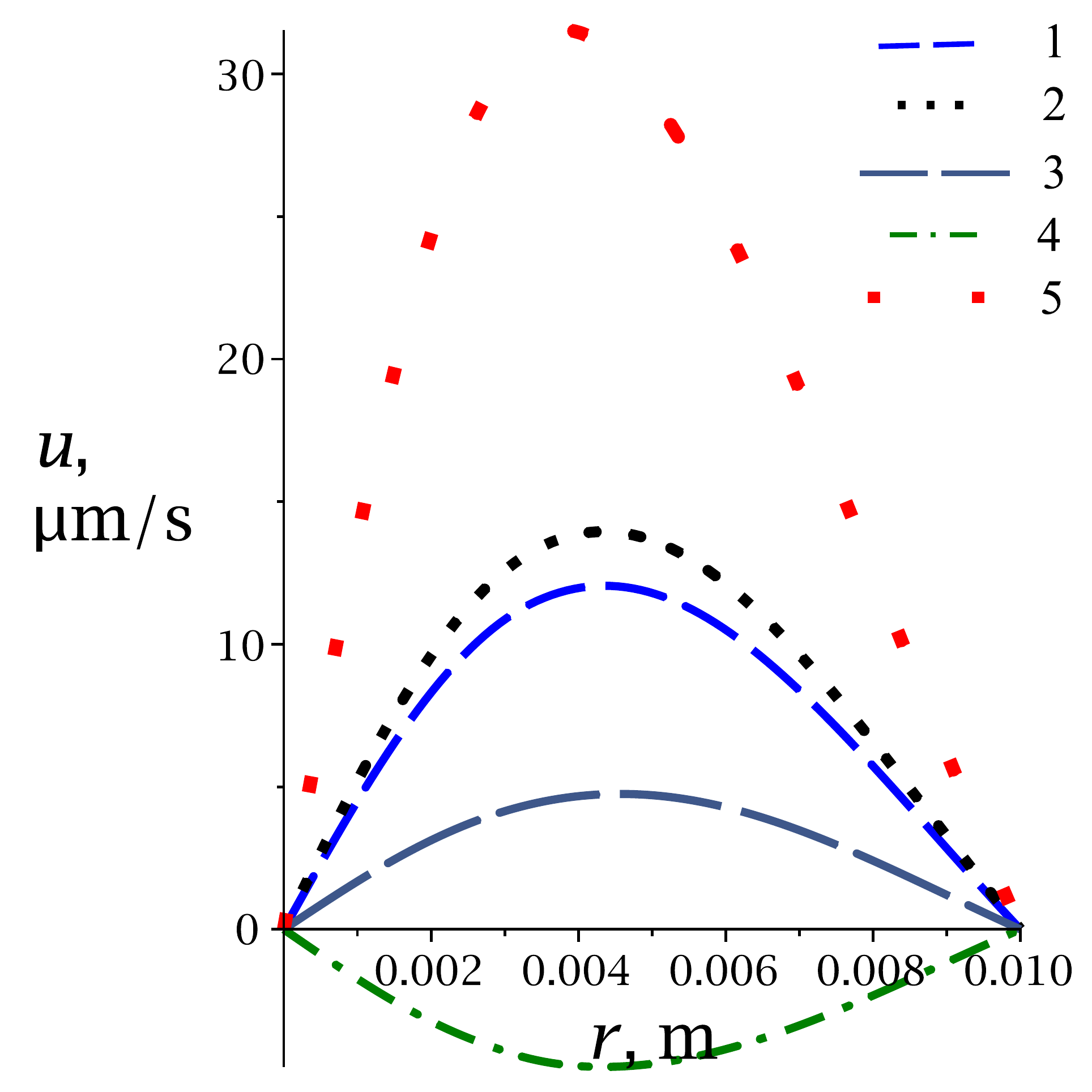}(b)
    \caption{Simulation results of (a) the spatiotemporal evolution of the fluid flow velocity averaged over the thickness of the liquid layer and (b) the spatial dependence of $u$ at time $t=25$~s for different values of the parameters: 1) $h_0=$ 700~$\mu$m, $q=$ 10$^4$~W/m$^2$; 2) $h_0=$ 700~$\mu$m, $q=$ 10$^4$~W/m$^2$, $\alpha_e=0$; 3) $h_0=$ 700~$\mu$m, $q=$ $0.5 \times 10^4$~W/m$^2$; 4) $h_0=$ 700~$\mu$m, $q=$ $-0.5 \times 10^4$~W/m$^2$; 5) $h_0=$ 400~$\mu$m, $q=$ $0.5 \times 10^4$~W/m$^2$.}
    \label{fig:heater_u}
\end{figure}

\begin{figure}[hbt]
    \includegraphics[width=0.65\linewidth]{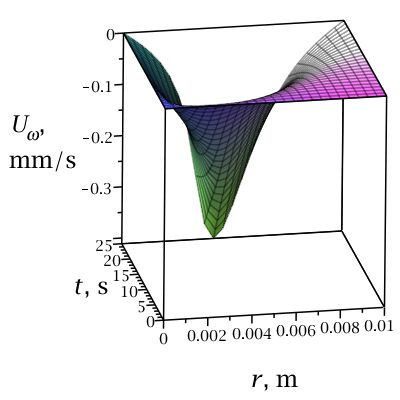}(a)\\
    \includegraphics[width=0.65\linewidth]{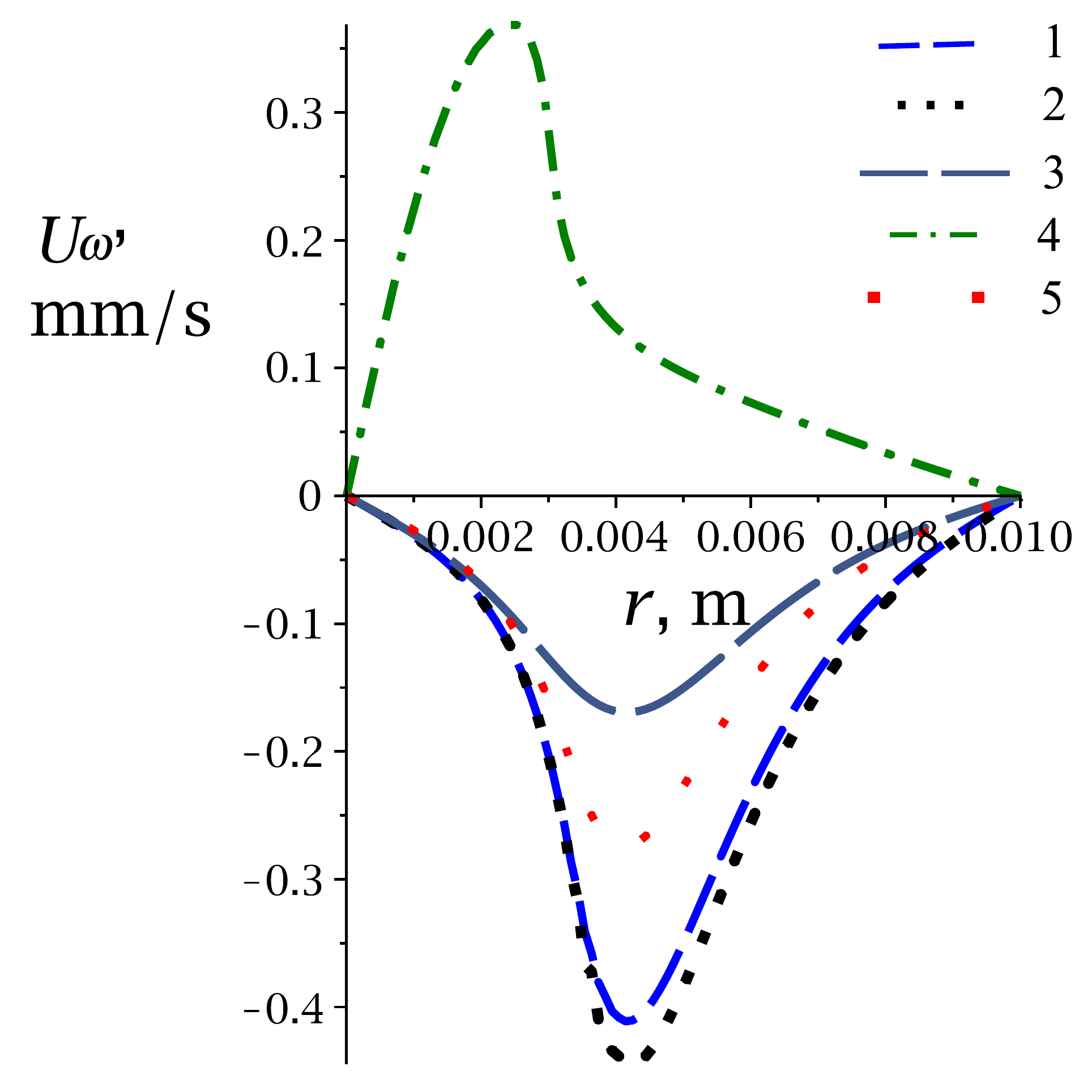}(b)
    \caption{Simulation results of (a) the spatiotemporal evolution of the fluid flow velocity near the substrate and (b) the spatial dependence of $U_\omega$ at time $t=25$~s for different values of the parameters: 1) $h_0=$ 700~$\mu$m, $q=$ 10$^4$~W/m$^2$; 2) $h_0=$ 700~$\mu$m, $q=$ 10$^4$~W/m$^2$, $\alpha_e=0$; 3) $h_0=$ 700~$\mu$m, $q=$ $0.5 \times 10^4$~W/m$^2$; 4) $h_0=$ 700~$\mu$m, $q=$ $-0.5 \times 10^4$~W/m$^2$; 5) $h_0=$ 400~$\mu$m, $q=$ $0.5 \times 10^4$~W/m$^2$.}
    \label{fig:heater_U_omega}
\end{figure}

The results of numerical calculations allow us to demonstrate the
evolution in space and time of the film thickness
(Fig.~\ref{fig:heater_h}a), the particle concentration
(Fig.~\ref{fig:heater_phi}a), the temperature of the liquid
(Fig.~\ref{fig:heater_Tl}a), the substrate
(Fig.~\ref{fig:heater_Ts}a), the average velocity
(Fig.~\ref{fig:heater_u}a), and the velocity of the liquid flow near
the substrate (Fig.~\ref{fig:heater_U_omega}a) at the initial stage
of the process. At the initial time, the free surface of the liquid
is flat. But, later in time, it begins to be curved. The film bends
down in the central area of the cell and rises slightly near the
wall (Fig.~\ref{fig:heater_h}a). This happens for two reasons. First,
in the central region, due to heating, the surface tension of the
liquid decreases.  Secondly, as the temperature increases, the
evaporation rate increases too. This precedes the thermocapillary
rupture of the film, which should occur later in time. The film
thickness $h$ decreases by about 13\% in the central region of the
cell during the heating time $t=25$~s. Note that this time refers to the period of intensive growth of the particle cluster (Fig.~\ref{fig:SquarePatternProfile}a). This intensive growth ends at a time of $t\approx$ 60~s when the cluster area is about 80\% of its maximum. Then there is a slow increase in the area up to the time $t\approx$ 150~s when the quasi-stationary mode is reached. After this moment, the cluster area practically does not change.

Simulation allows us to understand how key parameters affect the behavior of the system. Let us consider the effect of the heat flux density $q$, the initial film thickness $h_0$, and the presence or the absence ($\alpha_e=0$) of evaporation. When the heating is halved, the film thickness in the central part of the cell becomes approximately 0.05~mm higher at $t=25$~s (Fig.~\ref{fig:heater_h}b). In this case, the value of $h$ decreases by about 0.025~mm near the wall of the cell. Thus, a decrease in $q$ leads to a decrease in the curvature of the surface. The thermocapillary thinning of a liquid film becomes less pronounced due to the lower temperature gradient of the liquid (Fig.~\ref{fig:heater_Tl}b). At a negative value of $q$ (cooling mode), the film surface becomes convex because the liquid temperature gradient changes direction. The presence or absence of evaporation does not lead to significant differences in the shape of the film surface. But, even a slight difference in the curvature of the film surface can significantly affect the Laplace pressure gradient on which the capillary flow depends. At a lower value of the initial film thickness, the model predicts a deeper central thinning of the film (the surface curvature increases with a decrease in $h_0$) since the liquid warms up faster and the liquid temperature gradient is higher in this case. This leads to an increase in the surface tension gradient, which affects the shape of the meniscus.

The decrease in the liquid volume in the area of the heater and the mass transfer by the flow leads to an increase in the volume fraction of the particles $\phi$. In a relatively short period of time, the value of $\phi$ has increased by almost an order of magnitude compared to $\phi_0$ in the central area of the cell
(Fig.~\ref{fig:heater_phi}a). Peclet number
$\mathrm{Pe}_\phi=U_\mathrm{max} R_\mathrm{in}/D \approx 3\times 10^9$, where the characteristic velocity $U_\mathrm{max} \approx$ 1~mm/s according to experimental measurements (the maximum velocity $U_\omega$). At such value of solutal Peclet number, $\mathrm{Pe}_\phi \gg 1$, convective mass transfer prevails over diffusion. The diffusion term in the equation~\eqref{eq:convectionDiffusion} is taken into account solely for reasons of computational stability.

In the absence of evaporation, the volume fraction of particles in the center of the cell at time $t=$ 25~s is approximately 3\% higher (Fig.~\ref{fig:heater_phi}b). This may be because evaporation counteracts the transfer of solid matter toward the heater by affecting the fluid flow (Fig.~\ref{fig:heater_U_omega}b). A decrease in the heat flux density leads to a decrease in $\phi$ in the region of $r=0$ since the fluid flow velocity depends on $q$. In the cooling mode ($q<0$), the particle volume fraction decreases near the copper rod because the fluid flow changes direction (Fig.~\ref{fig:heater_u}b). In the case of $h_0=$ 700~$\mu$m, the model predicts particle motion to the wall of the cell at the cooling mode for the specified parameter $q$ value. This is qualitatively consistent with the result of the previous experiment for $h_0=$ 400~$\mu$m at a lower value of $q$~\cite{AlMuzaiqer2021}. A decrease in the initial film thickness from 700 to 400~$\mu$m at a constant $q$ leads to an increase in $\phi$ approximately twice larger, due to increased fluid flow.

The temperature of the liquid and the substrate increases over time, and this increase is particularly intense in the central part of the cell (Fig.~\ref{fig:heater_Tl}a, \ref{fig:heater_Ts}a). In the considered time interval, the temperature of the substrate $T_s$ in the area of the heating element has increased by 12~K. Instead, near the wall of the cell, the temperature of $T_s$ has not changed much.  This is due to the long relaxation time of the heat in the substrate,
$t_{rs} = \pi R_\mathrm{in}^2 / \chi_s \approx$ 819~s, where
the thermal diffusivity $\chi_s = k_s/ (c_s \rho_s)$
m$^2$/s. In addition, there is a constant transfer of heat from the bottom of the cell (substrate) to the liquid, since $T_s > T_l$ for any value of $r$ at $t>0$. Thermal Peclet number $\mathrm{Pe}_T=
u_\mathrm{max} R_\mathrm{in}/ \chi_l \approx$ 1.6, where the thermal diffusivity $\chi_l = k_l/ (c_l \rho_l)$ m$^2$/s.
The maximum value of the average velocity is taken from the results of our calculations, $u_\mathrm{max} \approx 10^{-5}$~m/s. This value of the Peclet number, $\mathrm{Pe}_T$, indicates that thermal diffusion and convection are equally important for heat distribution. During this time of heating ($t=25$~s), the temperature of the liquid in the central part of the cell increases by about 1.7~K. Near the wall, $T_l$ decreases by about 0.7~K. It is associated with evaporation since the molecules with the highest kinetic energy leave the liquid surface and pass into the vapor phase.  During this period of time, convection and thermal diffusion do not have time to compensate for this cooling yet. The temperature difference at the opposite boundaries of the calculated domain $\Delta T_l \approx $ 2.4~K. Thus, the appearance of the flow is affected not only by the deviation of the free surface of the film from the equilibrium shape but also by the gradient of the surface tension that occurs due to the nonuniform temperature $T_l$.

The model predicts that the evaporation leads to a decrease in the temperature of the liquid by about 1~K (Fig.~\ref{fig:heater_Tl}b). But, this does not have a noticeable effect on the temperature of the substrate (Fig.~\ref{fig:heater_Ts}b). A decrease in the value of $q$ leads to a decrease in the temperature of the substrate and the liquid. A negative value of the heat flux density changes the direction of the temperature gradient $T_s$ and $T_l$. The film thickness $h_0$ affects the temperature $T_l$ averaged over the thickness of the liquid layer. The thinner the liquid layer, the stronger and faster it warms up (Fig.~\ref{fig:heater_Tl}b). The model predicts that the temperature $T_s$ does not depend on the parameter $h_0$ (Fig.~\ref{fig:heater_Ts}b).

The radial flow velocity $u$ averaged over the thickness of the liquid layer increases from 0 to $10^{-5}$~m/s over the considered time interval. The positive velocity sign indicates that the fluid transfer occurs from the center toward the wall of the cell. This is due to the action of capillary forces and the change in the equilibrium shape of the film when a surface tension gradient occurs.  Due to the increase in temperature, the value of $\sigma$ decreases in the central part of the cell. Thus, the thermocapillary flow of liquid along the free surface will be directed from the center toward the wall of the cell, i.e., toward  the area of a higher value of the surface tension. Near the substrate, the liquid flow is directed toward the central region ($U_\omega < 0$), where it transfers the particles. The largest values of $u$ and $U_\omega$ in the plots (Fig.~\ref{fig:heater_u}a, \ref{fig:heater_U_omega}a) are observed in the region of $r \approx$ 4~mm, where a relatively sharp drop in temperature $T_l$ begins in the direction of the wall. The peak velocity is $U_\omega \approx $ -0.4~mm/s, which corresponds on order of magnitude to the experimental results. The opposite signs of the values $U_\omega$ and $u$ indicate that the flow near the free surface is more rapid than near the substrate.

A decrease in the heat flux density leads to a decrease in the modulus of the fluid flow velocity (Fig.~\ref{fig:heater_u}b, \ref{fig:heater_U_omega}b). The flow direction is inverted when the sign of $q$ changes. A decrease in $h_0$ at a constant $q$ leads to an increase in the modulus of the flow velocity due to the faster warming of the liquid layer. Evaporation contributes to the reduction of $|u|$ and $|U_\omega |$ by about 14~\% and 9~\%, respectively. Most likely, the evaporation increases the capillary flow, which counteracts the  Marangoni flow. But, at the same time, the thermocapillary flow prevails over the compensation flow. Nevertheless, evaporation has an impact on the system, so this controlled method of particle assembly into a cluster may be considered as one of the hybrid methods of evaporative lithography~\cite{Kolegov2020}. In the absence of evaporation, this method will be considered as thermocapillary particle assembly. The obtained numerical results are in qualitative agreement with the results of the performed experiments.

\section{Conclusion}
\label{conclusion}
Evaporative lithography is a promising field that is useful for microelectronics, nanotechnology, medicine, etc. Controlling the particle deposition and the structure formation with the required geometry and morphology on solid substrates is one of the key problems in technologies for creating materials and coatings with various functional properties. Here, we studied the process of the cluster formation of polystyrene microparticles distributed on a substrate under a layer of evaporating liquid (isopropanol), controlled by a point heater mounted in the center of the substrate.  Local heating creates a thermocapillary flow in the liquid layer, which is directed along the free surface from the heater toward the wall of the cell (from the region of low surface tension toward the region of high surface tension). The inward flow near the substrate, due to the viscous friction of the liquid, creates a force acting on the particles, which leads to the transfer of the particles to the heater area. As a result, the particles are accumulated in a cluster in the center of the cell. Experiments have shown that the size of the cluster at a constant power of the heat source depends on the thickness of the liquid layer. An increase in the layer thickness leads to a decrease in the area of the substrate occupied by the cluster. This indicates that a multilayer particle deposit is formed in the thick layer. Therefore, it is possible to control not only the cluster area but also the number of particle layers. Using PIV method, it has been established that the maximum values for the particle velocity are reached near the boundary of the growing cluster and increase with the temperature gradient. A mathematical model based on the lubrication approximation has been developed. The model allows us to describe the spatiotemporal evolution in the thickness of the liquid layer, the particle concentration, the temperature of the liquid and the substrate, as well as the flow velocity near the substrate. The obtained numerical results are in qualitative agreement with the experimental results. The model predicts that nonuniform evaporation slows down the thermocapillary flow that transfers the particles toward the heater along the substrate. This may be due to the competition between the capillary flow and the Marangoni flow. The work done allowed us to formulate the conclusion that the heat flow from the heater affects the geometry of the cluster for two reasons. First, the Marangoni flow velocity depends on the temperature gradient. Second, the decrease in film thickness near the heater depends on the temperature.

\begin{acknowledgments}
Part of the work related to mathematical modeling was supported by Grant No. 18-71-10061 from the Russian Science Foundation (K.S.K.). The experimental part of the research was supported by the Ministry of Science and Higher Education of the Russian Federation as part of World-class Research Center program: ``Advanced Digital Technologies'', Contract No. 075-15-2020-935 (N.A.I., V.M.F.). The experimental work of M.A.Al-M. was supported by Grant No. 19-31-90099 from the Russian Foundation for Basic Research.  The authors thank their colleague Fabio Grazioso for his careful proofreading of the manuscript.
\end{acknowledgments}

\section*{Data Availability Statement}

The data that support the findings of this study are available
from the corresponding author upon reasonable request.

\appendix

\section{Derivation of equations}\label{sec:appendix}

Here, we derive the heat transfer equation for a thin film. Let us consider the balance of thermal energy in the allocated
elementary volume $\Omega$ (Fig.~\ref{fig:domain}). The change in thermal energy in this volume over time $\Delta t$
$$
2\pi r\, \delta r\, \rho_l c_l \left( hT_l|_{r,t+\Delta t} - hT_l|_{r,t}
\right)
$$
is caused by a change in the mass of the volume itself
$$-2\pi r\, \delta r\, J\, \Delta t\, c_l\, T_l,$$
by the heat transfer due to flow
$$
-2\pi \left( rhuT_l|_{r+\delta r,t} - rhuT_l|_{r,t} \right)\rho_l
c_l\,\Delta t,
$$
by the cooling due to evaporation, heat exchange of the liquid and the substrate
$$
2\pi r\,\delta r\, \Delta t \left(  \alpha_{ls}\left( T_s - T_l \right) -
LJ\right),
$$
and by the diffusive heat transfer across the left and right boundaries of the domain
$\Omega$
$$
2\pi \left(  \left. rh\frac{\partial T_l}{\partial r}
\right|_{r+\delta r,t} - \left. rh\frac{\partial T_l}{\partial
    r}\right|_{r,t} \right)k_l\,\Delta t.
$$
Here, $\delta r$ is the distance between the left and right border of the domain $\Omega$. The entry like $\left. rh \,\partial T_l/ \partial r
\right|_{r+\delta r,t}$ should be taken as $(r+\delta r)\,h(r+\delta r,t)\,\partial T_l(r+\delta r,t)/\partial r$. Given the normal direction and assuming that the flow velocity coincides with the direction of the $r$ axis, we obtained the expression:
\begin{multline*}
    2\pi r\, \delta r\, \rho_l c_l \left( hT_l|_{r,t+\Delta t} - hT_l|_{r,t}
    \right)=\\-2\pi \left( rhuT_l|_{r+\delta r,t} - rhuT_l|_{r,t}
    \right)\rho_l c_l\,\Delta t +\\ 2\pi r\,\delta r\, \Delta t \left(
    \alpha_{ls}\left( T_s - T_l \right) - LJ\right) \\+ 2\pi \left(  \left.
    rh\frac{\partial T_l}{\partial r} \right|_{r+\delta r,t} - \left.
    rh\frac{\partial T_l}{\partial r}\right|_{r,t} \right)k_l\,\Delta t -2\pi r\, \delta r\, J\, \Delta t\, c_l\, T_l.
\end{multline*}
Then, we divided the left and right parts of this expression by $2\pi r\,\delta r\,
\Delta t\, \rho_l c_l$. In the limit transition ($\delta r \to 0$ and
$\Delta t \to 0$), we obtain equation~\eqref{eq:HeatTransferInLiquid}. We have obtained  equation~\eqref{eq:HeatTransferInSolid} using a similar procedure.

The model also needs to take into account the particle mass conservation law.
Change in the mass of particles in the volume $\Omega$ over time $\Delta t$ ($\rho_p=\rho_l$),
$$2\pi
r\, \delta r \left( \left. \phi h \right|_{r,t+\Delta t} - \left.
\phi h \right|_{r,t} \right) \rho_l,
$$
is caused by the transfer of particles by the flow,
$$-2\pi z_\omega \left(  \left. r U_\omega \phi_\omega \right|_{r+\delta r,t} -
\left. r U_\omega \phi_\omega \right|_{r,t} \right) \rho_l \Delta t,
$$
and by particle diffusion (taking into account Fick's law),
$$ 2\pi z_\omega \left(
\left. r \frac{\partial \phi_\omega}{\partial r} \right|_{r+\delta
    r,t} - \left. r \frac{\partial \phi_\omega}{\partial r}
\right|_{r,t} \right) D \rho_l \Delta t.$$ With $\phi_\omega =
\phi h/z_\omega$, we obtain the expression
\begin{multline*}
    2\pi r\, \delta r \left( \left. \phi h \right|_{r,t+\Delta t} -
    \left. \phi h \right|_{r,t} \right) \rho_l = \\ 2\pi \left( \left. r
    \frac{\partial h \phi}{\partial r} \right|_{r+\delta r,t} - \left. r
    \frac{\partial h \phi}{\partial r} \right|_{r,t} \right) D \rho_l
    \Delta t \\ -2\pi \left(  \left. r U_\omega h \phi \right|_{r+\delta
        r,t} - \left. r U_\omega h \phi \right|_{r,t} \right) \rho_l \Delta t.
\end{multline*}
Then, we divide the left and right parts of the resulting expression by $2\pi r
\delta r \rho_l \Delta t$ and, in the limit transition, we obtain  equation~\eqref{eq:convectionDiffusion}.

Also, if we consider mass conservation law for the suspension as a whole, the change in the mass of suspension in the volume $\Omega$ over time $\Delta t$
$$2\pi r\, \delta r \left( h(r,t+\Delta t) - h(r,t) \right) \rho_l,$$
is caused by the fluid flow,
$$-2\pi \left( \left. rhu(1-\phi) \right|_{r+\delta r,t} - \left. rhu(1-\phi) \right|_{r,t} \right)\rho_l \Delta t,$$
by the particle redistribution,
$$-2\pi z_\omega \left( \left. r U_\omega \phi_\omega \right|_{r+\delta r, t} -
\left. r U_\omega \phi_\omega \right|_{r, t} \right) \rho_l \Delta t,$$
and by the evaporation,
$$-2\pi r \, \delta r J(r,t)\, \Delta t.$$
By taking into account $\phi_\omega = \phi h/z_\omega$, we obtain the expression
\begin{multline*}
    2\pi r\, \delta r \left( h(r,t+\Delta t) - h(r,t) \right) \rho_l =\\
    -2\pi \left( \left. rhu(1-\phi) \right|_{r+\delta r,t} - \left.
    rhu(1-\phi) \right|_{r,t} \right)\rho_l \Delta t\\
    -2\pi \left( \left. r U_\omega h \phi \right|_{r+\delta r, t} -
    \left. r U_\omega h \phi \right|_{r, t} \right) \rho_l \Delta t -2\pi
    r \, \delta r J(r,t)\, \Delta t.
\end{multline*}
If we devide both parts of it by $2\pi r\, \delta r \rho_l \Delta t$, and in the limit transition, we obtain  equation~\eqref{eq:conservationMassLaw}.

\nocite{*}
\bibliography{aipsamp}

\end{document}